\RequirePackage{fix-cm}
\RequirePackage{multirow}
\documentclass[smallextended]{svjour3}    
\PassOptionsToPackage{numbers}{natbib}
\usepackage{natbib}
\usepackage{url}
\usepackage{graphicx}
\usepackage{array}
\usepackage{multirow}
\usepackage{verbatim}
\usepackage{hyphenat}
\usepackage{float}
\usepackage{tcolorbox}
\usepackage{enumitem}
\usepackage{balance}
\usepackage{todonotes}
\usepackage{threeparttable}
\usepackage{listings}
\usepackage{booktabs}
\usepackage{capt-of}

\usepackage{todonotes}
\usepackage{footmisc}

\definecolor{gray50}{gray}{.5}
\definecolor{gray40}{gray}{.6}
\definecolor{gray30}{gray}{.7} 
\definecolor{gray20}{gray}{.8}
\definecolor{gray10}{gray}{.9}
\definecolor{gray05}{gray}{.95}

\newlength\Linewidth
\def\findlength{\setlength\Linewidth\linewidth
	\addtolength\Linewidth{-4\fboxrule}
	\addtolength\Linewidth{-3\fboxsep}
}
\newenvironment{rqbox}{\par\begingroup
	\setlength{\fboxsep}{5pt}\findlength
	\setbox0=\vbox\bgroup\noindent
	\hsize=0.95\linewidth
	\begin{minipage}{0.95\linewidth}\normalsize}
	{\end{minipage}\egroup
	\textcolor{gray20}{\fboxsep1.5pt\fbox
		{\fboxsep5pt\colorbox{gray10}{\normalcolor\box0}}}
	\endgroup\par\noindent
	\normalcolor\ignorespacesafterend}

\newcommand{\rb}[1]{
	
	\begin{tcolorbox}[colback=gray!05,%gray background
		colframe=black,% black frame colour
		width=\columnwidth,% Use 5cm total width,
		arc=3mm, auto outer arc,
		boxrule=0.5pt,
		]
		#1
	\end{tcolorbox}
}

\newcounter{Finding}
\stepcounter{Finding}

\newcommand{\roundedbox}[1]{
	\rb{
		\noindent
		\textit{\textbf{Finding \theFinding}. #1}
	}
	\stepcounter{Finding}
}

\begin{document}
\title{Fault Prediction based on Software Metrics and SonarQube Rules. Machine or Deep Learning?}

\author{Francesco Lomio \and Sergio Moreschini \and Valentina Lenarduzzi}

\institute{
            Francesco Lomio and Sergio Moreschini \at
            Tampere University, Finland \\
            \email{francesco.lomio@tuni.fi}  
            \email{sergio.moreschini@tuni.fi}  
            \and
            Valentina Lenarduzzi \at
           LUT University, Finland \\
           \email{valentina.lenarduzzi@lut.fi}    
}

\date{Received: date / Accepted: date}

\maketitle

\begin{abstract}
\textit{Background.} Developers spend more time fixing bugs and refactoring the code to increase the maintainability than developing new features. Researchers investigated the code quality impact on fault-proneness focusing on code smells and code metrics. \\
\textit{Objective.}  We aim at advancing fault-inducing commit prediction based on SonarQube considering the contribution provided by each rule and metric.\\
\textit{Method.} We designed and conducted a case study among 33 Java projects analyzed with SonarQube and SZZ to identify fault-inducing and fault-fixing commits. Moreover, we investigated fault-proneness of each SonarQube rule and metric using Machine and Deep Learning models.  \\
\textit{Results.} We analyzed 77,932 commits that contain 40,890 faults and infected by more than 174 SonarQube rules violated 1,9M times, on which there was calculated 24 software metrics available by the tool. Compared to machine learning models, deep learning provide a more accurate fault detection accuracy and allowed us to accurately identify the fault-prediction power of each SonarQube rule. As a result, fourteen of the 174 violated rules has an importance higher than 1\% and account for 30\% of the total fault-proneness importance, while the fault proneness of the remaining 165 rules is negligible. \\
\textit{Conclusion.} Future works might consider the adoption of timeseries analysis and anomaly detection techniques to better and more accurately detect the rules that impact fault-proneness.

\end{abstract}

\section{Introduction}
\label{sec:Intro}
Software teams spend a significant amount of time trying to locate defects and fixing bugs~\cite{Zeller2009}. 
Fixing a bug involves isolating the part of the code that causes an unexpected behavior of the program and changing it to correct the error~\cite{Beller2018}. Bug fixing is a challenging task, and developers often spend more time fixing bugs and making the code more maintainable than developing new features~\cite{Murphy-Hill2015, Pan2009}. 

Different works addressed this problem~\cite{DAmbros2010, Haidar2017}, relying on different information, such as process metrics~\cite{Nagappan2005, Moser2008, Hassan2009}(number of changes, recent activity), code metrics~\cite{Subramanyam2003, Gyimothy2005, Nagappan2006} (lines of code, complexity) or previous faults~\cite{Ostrand2005, Hassan2005, Kim2007}. The research community also considered the impact of different code quality issues on fault-proneness, with a special focus on Fowler's code smells~\cite{Palomba2018, Gatrell2015, DAmbros2010, Saboury2017, LenarduzziJSS2020}.  

In our previous works we investigated the fault-proneness of SonarQube rules first with machine learning techniques~\cite{LenarduzziSANER2020} and second with classical statistic techniques~\cite{LenarduzziJSS2020}.
As a result, both works confirm that SonarQube rules, all together, provide a good prediction power of faults, even if SonarQube classifies most of the rules as non fault-inducing. 
However, the techniques adopted in our previous work, did not allow to identify the impact of each individual SonarQube rule on fault-proneness.  
As a result, developers commonly struggle to understand which metric or SonarQube rules they should consider to decrease the fault-proneness of their code~\cite{Vassallo2018}, especially because the ruleset includes more than 500 rules per development language.

SonarQube is one of the most frequently used open-source static analysis tools~\cite{LenarduzziSEDA2019, Avgeriou2020}, having been adopted by more than 120K users\footnote{https://www.sonarqube.org}, including more than 200K development teams and adopted by more than 100K open source projects\footnote{\label{sonarcloud}https://sonarcloud.io/explore/projects}. SonarQube analyze the code compliance against a set of rules classified as: \textit{Code Smells}, i.e.,  issues that increase change-proneness and the related maintenance effort;  \textit{Bugs}, i.e.,  issues that will result in a fault; and \textit{Security Vulnerabilities}\footnote{\label{sq-rules}SonarQube Rules: https://docs.sonarqube.org/display/SONAR/Rules}. SonarQube also assigns one of the following five \textit{severity} levels to each rule:  \textit{Blocker}, \textit{Critical}, \textit{Major}, \textit{Minor}, and \textit{Info}. Rules with a higher severity level have a higher impact on the system. As an example, a Sonar issue labeled as of severity \textit{Blocker} and type \textit{Bug} highlights a piece of code that \textit{''has a high probability to cause the application to crash or to corrupt the stored data''}\textsuperscript{\ref{sq-rules}}.

Similarly to our previous works \cite{LenarduzziSANER2020, LenarduzziJSS2020}, this paper investigates the relationship existing between the occurrence of SonarQube rules violations in software projects and  fault-proneness. Specifically, while previous work shows a significant correlation between the violation of a SonarQube rule and code change/fault-proneness, the empirical evidence provided so far is still limited because of:
\begin{itemize}
    \item Lack of analysis on the magnitude of the fault-proneness of each individual SonarQube rule. Because of the lack of a large cluster with modern GPUs, our previous work did not allow us to compute the fault-proneness of each SonarQube rule, but, instead, investigated the fault-proneness of all the SonarQube rules together.  
    \item Lack of comparison of the fault prediction power of software metrics and SonarQube rules. 
\end{itemize}

To cope with the aforementioned issues, in this paper we aim at advancing the fault-inducing commit prediction based on SonarQube rules and metrics.

Starting from the results obtained in our previous work~\cite{LenarduzziJSS2020}, we designed and conducted a case study among \textit{33 Java projects} of the Technical Debt dataset~\cite{LenarduzziPromise2019} analyzed with SonarQube version 7.5 that violated more than \textit{1,9M  of SonarQube rules}, and where the faults were determined applying the SZZ algorithm~\cite{SZZ}. 
We compared the fault prediction power of each SonarQube rule and software metrics detected by SonarQube using the three most accurate machine learning models identified in our previous work~\cite{LenarduzziJSS2020} and two deep learning models.

Results show that Deep learning models provide a more accurate fault detection accuracy compared with machine learning ones based on SonarQube rules. Moreover, deep learning models allow a clear identification of features set more informative to the fault identification, while machine learning models do not provide a clear distinction. Considering the metrics calculated by SonarQube, none of the models provide a good level of accuracy and, consequently, the features set identification. 

The contribution of this paper is 3-fold: 
\begin{itemize}
    \item An accurate identification of the fault-proneness of each Java rule detected by SonarQube
    \item A comparison of the prediction power of the fault-proneness of  SonarQube rules and metrics
    \item A comparison of the effectiveness and accuracy of machine learning and deep-learning techniques for the identification of fault-inducing rules and metrics 
\end{itemize}

% Looking at the accuracy metrics adopted in this work, we can noticed a consistent difference in magnitude for the individual accuracy metrics. Future works might consider the adoption of time series analysis and anomaly detection techniques to better and more accurately detect the rules that impact fault-proneness.

The remainder of this paper is structured as follows. In Section~\ref{sec:Background} we introduce the background in this work, introducing the original study, SonarQube violations and the different machine and deep learning models. Section~\ref{sec:CS}, describes the case study design, while Section~\ref{sec:Results} presents the obtained results. Section~\ref{sec:Discussion} discusses the results, and Section~\ref{sec:Threat} identifies threats to validity.
Section~\ref{sec:RW} describes the related works, while Section~\ref{sec:Conclusions} drawn the conclusion highlighting the future works. 
\section{Background}
\label{sec:Background}
In this Section, we illustrate the background of this work, introducing our previous study (called ``original''), SonarQube static analysis tool, and the Machine and Deep Learning models adopted in this study.

\subsection{The Original Study}
\label{sec:OriginalStudy}
In this Section, we illustrate the original study~\cite{LenarduzziSANER2020} and the obtained results. Moreover, we explain the reasons why we conducted this study, and we compare it with the original one. We followed the guidelines proposed by Carver for reporting replications~\cite{CarverReplication}.

We decided to consider for this study, only the paper~\cite{LenarduzziSANER2020} since – as far as we know – this is the only one that provide a ranking of importance of SonarQube issues that could induce bugs in the source code.

Moreover, two of the authors of this paper are also authors of the original study. 

The original study investigated the fault-proneness of SonarQube rules in order to understand if rules classified as ''Bug'' are more fault-prone than security and maintainability rules (``vulnerability'' and ``code smell''). Moreover, the original study evaluated the accuracy of the SonarQube quality model for the bugs prediction. 
As context, the original study analyzed 21 randomly selected mature Java projects from the Apache Software Foundation. All the commits of the projects were analyzed with SonarQube (version 6.4), and the commits that induced a fault were determined applying the SZZ algorithm~\cite{SZZ}.
The SonarQube rules fault proneness was investigated with seven Machine Learning algorithms (Decision Trees \cite{Breiman1984ClassificationTrees}, Random Forest \cite{breiman2001random}, Bagging \cite{Breiman1996BaggingPredictors}, Extra Trees \cite{Geurts2006ExtremelyTrees}, Ada Boost \cite{Freund1997ABoosting}, Gradient Boost \cite{FriedmanGreedyMachine}, XG Boost \cite{Chen2016XGBoost:Systemb}). 

Results show that only a limited number of SonarQube rules can be considered really fault-prone. 

Differently from the original study we considered the \textit{33 Java projects} of the Technical Debt dataset~\cite{LenarduzziPromise2019}, analyzed with SonarQube version 7.5 that are infected by more than \textit{1,9M SonarQube rules} violated, on which there was calculated \textit{24 software metrics}, and where the faults are determined applying the SZZ algorithm~\cite{SZZ}. 
Moreover, we adopted \textit{deep learning models} and we made a comparison between the detection accuracy of deep learning and machine learning models in order to identify which ones better predict a fault and identify the feature selection importance. We adopted the three machine learning models that exhibit the best accuracy performance (AUC = 80\%) in the original study.

\begin{table}[H]
\caption{Study Design Comparison}
\label{tab:DesignComparison}
\centering
\begin{tabular}{l|l|l} \hline 
& \textbf{Original Study}~\cite{LenarduzziSANER2020} & \textbf{New Study} \\ \hline 
\#Projects & 21 & 33\\
\#Commits & 39,518 & 77,932\\
SonarQube tool version & 6.4 & 7.5 \\
SonarQube rules & 231,453 & 1,941,508  \\
Faults & 4,505 & 40,890 \\
Software metrics & 0 & 33 \\
Machine Learning models & 8 & 3*\\
Deep Learning models & 0 & 2 \\
\hline 
\multicolumn{3}{l}{*the best ones among the 8 adopted in~\cite{LenarduzziSANER2020}} \\ 
\end{tabular}
\end{table}

\subsection{SonarQube}
\label{sec:SonarQube}
SonarQube is one of the most common open-source static code analysis tools adopted both in academia~\cite{LenarduzziICSE2017, LenarduzziSEDA2018} and in industry~\cite{VassalloEMSE19}. SonarQube is provided as a service from the sonarcloud.io platform or it can be downloaded and executed on a private server.

SonarQube calculates several metrics such as the number of lines of code and the code complexity, and verifies the code's compliance against a specific set of ``coding rule'' defined for most common development languages. 
In case the analyzed source code violates a coding rule or if a rule is outside a predefined threshold, SonarQube generates an ``issue''. 
SonarQube includes Reliability, Maintainability and Security rules. 

Reliability rules, also named ``bugs'' create issues (code violations) that ``represents something wrong in the code'' and that will soon be reflected in a bug.  ``Code smells'' are considered  ``maintainability-related issues'' in the code that decreases code readability and code modifiability. It is important to note that the term ``code smells'' adopted in SonarQube does not refer to the commonly known code smells defined by Fowler et al.~\cite{Fowler1999} but to a  different set of rules. Fowler et al.~\cite{Fowler1999} consider code smells as 
``surface indication that usually corresponds to a deeper problem in the system'' but they can be indicators of different problems (e.g., bugs,  maintenance effort, and code readability) while rules classified by SonarQube as 
``Code Smells'' are only referred to maintenance issues. Moreover,  only four of the 22 smells proposed by Fowler et al. are included in the rules  classified as 
``Code Smells'' by SonarQube (Duplicated Code, Long Method, Large Class, and Long Parameter List).   

SonarQube also  classifies the rules into five \textit{severity} levels\footnote{SonarQube Issues and Rules Severity:'  https://docs.sonarqube.org/display/SONAR/Issues}: Blocker, Critical, Major, Minor, and Info. 
% \todo[inline]{Sergio: non dovrebbe essere SQ-violations? quando scriviamo SonarQube SQ sono sempre maiuscole addirittura nella dicitura ufficiale quindi mi sorge spontaneo pensare che sia SQ maiuscolo}

In this work, we focus on the SonarQube violations, which are reliability rules classified  as ``bugs'' by SonarQube, as we are interested in understanding whether they are related to faults. Moreover, we consider the 33 software metrics calculated by SonarQube. 
% \todo[inline]{add list of SQ metrics and describe them}
In the replication package (Section~\ref{sec:Replicability}) we report all the violations present in our dataset. In the remainder of this paper, column ``\textit{squid}'' represents the original rule-id (SonarQube ID) defined by SonarQube. We did not rename it, to ease the replicability of this work.  In the remainder of this work, we will refer to the different SonarQube violations with their id (squid). The complete list of violations can be found in the file ``SonarQube-rules.xsls'' in the online raw data.

\subsection{Machine Learning models}
\label{sec:MachineLearning}
We selected three machine learning models (\textit{Gradient Boost}\cite{FriedmanGreedyMachine}, \textit{Random Forest~\cite{breiman2001random}}, and \textit{XGboost}~\cite{Chen2016XGBoost:Systemb}) that they turned out to be the most accurate in the faults prediction in our original study~\cite{LenarduzziSANER2020}. 
As for~\cite{LenarduzziSANER2020}, Gradient Boosting and Random Forest are implemented using the library \textit{Scikit-Learn}\footnote{https://scikit-learn.org} with their default parameters. XGBoost model is implemented using the \textit{XGBoost} library\footnote{https://xgboost.readthedocs.io}. All the classifiers are fitted using 100 estimators.

\vspace{2mm}
\textbf{Random Forest.} Random Forest~\cite{breiman2001random} is an ensemble technique based on decision trees. The term ensemble indicates it uses a set of "weak" classifiers that help solving the assigned task. In this specific case, the week classifiers are multiple decision trees.

Using a randomly chosen subset of the original dataset, an arbitrary amount of decision trees is generated~\cite{Breiman1996BaggingPredictors}. In the case of random forest, the subset is created with replacement, meaning that a sample can appear multiple times. Moreover, it is also chosen a subset of the features of the original dataset, without replacement (appear only once). This helps reducing the correlation between the individual decision tress.
With this setup, each tree trained on a specific subset of the data, and it can make prediction on unseen data. The results of the predictions of all the decision trees used are averaged and used by the random forest to classify the input.

The process of averaging the prediction of multiple decision trees, allows the random forest classifier to better generalize the data and overcome the overfitting problem to which decision trees are prone. Also, using a randomly selected subset of the original dataset, the individual trees are not correlated between one another.
This is specifically important in our case, as in this study we are using a high number of features, and therefore the probability of the feature being correlated to one another, increases.

\vspace{2mm}
\textbf{Gradient Boosting.} Gradient Boosting~\cite{FriedmanGreedyMachine} is another ensemble model which, compared to the random forest, generates the individual weak classifiers sequentially during the training process. In this case as well we are using a series of decision trees as weak classifiers. The gradient boosting model, creates and trains at first only one decision tree. After each iteration, another tree is grown in order to improve the accuracy of the model and therefore minimize the loss function. This process continues until a predefined number of decision trees has been created, or the loss function no longer improves.

\vspace{2mm}
\textbf{XGBoost.} The last classical model used, is the XGBoost~\cite{Chen2016XGBoost:Systemb}. This is nothing but a better performing implementation of the Gradient Boosting algorithm. It allows for faster computation and parallelization compared to the gradient boosting. It can therefore result in better computational and overall performance compared to the latter, and can be more easily scaled for the use with high dimensional data, as it is the one we are using.

\subsection{Deep Learning models}
\label{sec:DeepLearning}
Deep learning is a subset of machine learning (ML), based on the use of artificial neural network. The term deep, indicates the use of multiple layers in the neural network architecture: the classical artificial neural network is the multilayer perceptron (MLP), which comprises an input layer, and output and a hidden layer in between. This structure put the limit to the quantity of information that the network can learn and use for its task. By adding more layers, it allows the network to increase the amount of information that the network is able to extract from the raw input, improving its performance.

While machine learning models become progressively better at whatever their function is, they still need some guidance, especially in the way the features are provided in input. In most of the cases it is necessary to perform some basic to advance feature engineering before being able to feed them to the model for training.
Deep learning models, on the other hand, thanks to their ability to progressively extract higher level features from the input in the multiple layers of their architecture, require little to no previous feature engineering. This is particularly helpful when dealing with high dimensional data.

Also, as seen section~\ref{sec:MachineLearning}, most of the classical machine learning models suffer in performance when dealing with large datasets and high dimensional data. Deep learning models, on other hands, can be helpful as thanks to the different type of architectures they can be more scalable and flexible.

In this Section, we briefly introduce the Deep Learning-based techniques we adopted in this work: \textit{Fully Convolutional Network}(FCNN)~\cite{wang2017time} and \textit{Residual Network}(ResNet)~\cite{wang2017time}. 

% \todo[inline]{@francesco: PERCHE QUESTE DUE?}
% \todo[inline]{aggiungere reference a questa parte ( e controllare anche reference ML)}
These two approaches are adopted from \cite{fawaz2019deep}, where it was shown that their performance is superior to multiple other methods tested. In particular, \textit{Fawaz et al.} showed in their work that the FCNN and the ResNet were the best performing classifiers in the context of the multivariate time series classification. This conclusion was obtained testing 9 different deep learning classifiers on 12 multivariate time series datasets.
% \todo[inline]{Sergio: Secondo me è essenziale una votla inserite le references ai paper ML prendere le figure delle reti (citando la fonte) di modo da rendere piu diretto l'approccio.}

% \todo[inline]{Vale: FAI!}

% \todo[inline]{FRA: Sarebbe furbo farlo, hai ragione. Il problema è che le reti sono particolarmente grandi e i grafici di queste due specifiche non credo si trovino. In ogni caso prova a guardare sulla  \cite{fawaz2019deep}. L'altro problema è che alcune dimensioni sono diverse perché adattate. Al limite se riesci metti le figure, e se non capisci le cosa modificare dimmelo e faccio io domenica}

\vspace{2mm}
\textbf{Residual Network.} The first deep learning model used is a residual network (ResNet)~\cite{wang2017time}. Among the many different type of ResNet developed, the one we used is composed by 11 layers of which 9 are convolutional. Between the convolutional layers it has some shortcut connection which allows the network to learn the residual~\cite{he2016deep}. In this way the network can be trained more easily, as there is a direct flow of the gradient through the connections. Also, the connection helps in reducing the \textit{vanishing gradient effect}, which prevent deeper neural network to properly train.

In this work we used the ResNet shown in \cite{fawaz2019deep}. It consists of 3 residual blocks, each composed of three 1-dimensional convolutional layers alternated to pooling layers, and their output is added to input of the residual block. The last residual block, is followed by a global average pooling (GAP) layer~\cite{lin2013network} instead of the more traditional fully connected layer. The GAP layer allows the features maps of the convolutional layers to be recognised as a category confidence map. Moreover, it reduces the number of parameters to train in the network, making it more lightweight, and reducing the risk of overfitting, when compared to the fully connected layer.

\vspace{2mm}
\textbf{Fully Convolutional Neural Network.} The second method used, is a fully convolutional neural network (FCNN)~\cite{wang2017time}. This network, compared to the ResNet, does not present any pooling layer (hence the name), which keeps the dimension of the time series remains unchanged throughout the convolutions. As for the ResNet, after the convolutions, the features are passed to a global average pooling (GAP) layer.

The FCNN used in this work is adopted from \cite{fawaz2019deep}. This implementaion consists of 3 convolutional blocks, each composed by a 1-dimensional convolution and by a batch normalization layer~\cite{ioffe2015batch}. It uses a rectified linear unit (ReLU)~\cite{nair2010rectified} activation function. The output of the last convolutional block are fed to the GAP layer, fully connected to a traditional softmax for the time series classification.

\section{Case Study Design}
\label{sec:CS}

We designed our empirical study as a case study based on the guidelines defined by Runeson and Host~\cite{Runeson2009} and follows the ACM/SIGSOFT Empirical Standards~\cite{ralph2020acm}.
In this Section, we describe the empirical study including the goal and the research questions, the study context, the data collection and the data analysis. 

\subsection{Goal and Research Questions}

The goal of this paper is to identify the fault-proneness of software metrics and rules detected by SonarQube.
Based on the aforementioned goal, we derived the following Research Questions (RQ$_s$).

\vspace{2mm}
\begin{tabular}
{@{}p{0.8cm}p{10cm}@{}}
\textbf{RQ$_1$} & What is the fault proneness of all the SonarQube rules?\\
\textbf{RQ$_{1.1}$} & What is the fault proneness of each individual SonarQube rule?\\
 
\textbf{RQ$_2$} & What is the fault proneness of all the metrics calculated by SonarQube?\\
\textbf{RQ$_{2.1}$} & What is the fault proneness of each metric calculated by SonarQube?\\

%   \textbf{RQ$_{1.1}$} & What is the fault proneness of the SonarQube rules and metrics, when using Machine Learning models?\\
% \textbf{RQ$_{1.2}$} & What is the fault proneness of the SonarQube rules and metrics, when using Deep Learning models?\\
% \textbf{RQ$_2$} & What is the fault proneness of each software metric calculated by SonarQube?\\

% \textbf{RQ$_2$} & Does Deep Leaning models enable a more accurate SonarQube rules and metrics fault proneness identification when compared to a machine learning models? \\
\end{tabular}

\vspace{2mm}
More specifically, in \textbf{RQ$_1$} we aim at investigating the impact of all the SonarQube rules on fault-proneness. The goal is to understand how accurate the prediction can be for fault-proneness if developers do not violate all the SonarQube rules. 
However, as reported by Vassallo et al.~\cite{Vassallo2018}, developers commonly get confused by the large number of rules, especially because their severity assigned by SonarQube is not actually correlated with the fault proneness~\cite{LenarduzziJSS2020}\cite{LenarduzziSANER2020}. Therefore, in order to help developers and SonarQube producers to better understand the actual proneness of each rule, in  \textbf{RQ$_{1.1}$ } we aim at investigating the impact of each rule individually on the fault-proneness independently from its type (Bugs, Code Smells, and Vulnerabilities) . The ultimate goal is to recommend a reduced set of rules that developers should consider instead of the complete set of more than 500 rules. 

Software metrics have been considered as good predictors for fault-proneness for several decades~\cite{DAmbros2010}. Therefore, in \textbf{RQ$_2$} we are interested to investigate the fault proneness of the 24 metrics detected by SonarQube. In order to have a baseline for the next RQ, in this RQ we aim at investigating the impact of all the metrics on fault proneness. 
If the magnitude of the phenomenon is small—i.e., all the software metrics  calculated by SonarQube cannot be used to predict fault-prone commits, then studying their individual  impact on fault-proneness  might not be worthwhile.
In \textbf{RQ$_{2.1}$},  we aim at investigating the impact of each metric individually on fault-proneness, so as to understand if also in this case it is possible to consider only  a limited set of metrics.

% In \textbf{RQ$_2$}, we perform a comparison between the detection accuracy of deep learning and machine learning techniques in order to identify which ones better predict a fault-inducing commit and identify the feature selection importance. This results is important to provide to the developers a more accurate approach for the fault-inducing commits detection. 

% As describe in Section~\ref{sec:Background}, Deep Learning models can better fits our data, so we hypothesized that the fault inducing commits prediction can be more accurate in term of models accuracy metrics (e.g. AUC) and feature importance selection (e.g. SonarQube rules or software metrics). 

\subsection{Study Context}
\label{Context}
As context, we considered the projects included in the Technical Debt Dataset~\cite{LenarduzziPromise2019}. The data set contains 
33 Java projects from the Apache Software Foundation (ASF) repository\footnote{http://apache.org}. The projects in the data set were selected  based on ``criterion sampling''~\cite{Patton2002}, that fulfill all of the following criteria: developed in Java, older than three years, more than 500 commits and 100 classes, and usage of an issue tracking system with at least 100 issues reported. 
The projects were selected also maximizing their diversity and representation by considering a comparable number of projects with respect to project age, size, and domain. Moreover, the 33 projects can be considered mature, due to the strict review and inclusion process required by the ASF. Moreover, the included projects regularly review their code and follow a strict quality process\footnote{https://incubator.apache.org/policy/process.html}. More details on the data set can be found in~\cite{LenarduzziPromise2019}.

For each project, Table~\ref{tab:SelectedProjects} reports the number of commits analyzed, the number of faults detected, and the number of Sonarqube rules violated. 

\begin{table} []
\footnotesize
\centering
\caption{The selected projects} 
\label{tab:SelectedProjects} 
\begin{tabular}
{@{}l|r|r|r@{}}
% {@{}p{3.2cm}|p{1.6cm}|p{1.2cm}|p{3cm}@{}}
\hline 
\textbf{Project}	&	\textbf{\#Commits}	&	\textbf{\#Faults}	&	\textbf{\#SQ rule violations}	\\\hline 
Accumulo   	&	2,641	&	2,250	&	1,429,757	\\
Ambari	&	13,397	&	17,722	&	41,612	\\
Atlas   	&	2,336	&	1,990	&	35,776	\\
Aurora	&	4,012	&	628	&	7526	\\
Batik 	&	2,097	&	1,160	&	31,691	\\
Beam 	&	2,865	&	1,723	&	8,449	\\
Bcel 	&	10,210	&	3,218	&	85,018	\\
Beanutils 	&	1,324	&	242	&	5,182	\\
Cli 	&	1,192	&	346	&	37,408	\\
Codec	&	896	&	182	&	58,073	\\
Cocoon 	&	1,726	&	327	&	2,041	\\
Collections 	&	2,982	&	135	&	11,118	\\
Configuration 	&	2,895	&	73	&	5,612	\\
Deamon 	&	980	&	190	&	393	\\
Dbcp 	&	1,861	&	284	&	3,696	\\
Dbutils 	&	645	&	159	&	644	\\
Digester  	&	2,145	&	149	&	4,947	\\
Exec 	&	617	&	444	&	762	\\
Felix 	&	596	&	147	&	11,340	\\
FileUpload 	&	922	&	282	&	769	\\
Httpcomponents Client   	&	2,867	&	463	&	10,803	\\
HttpComponents Core 	&	1,941	&	188	&	9,531	\\
Io 	&	2,118	&	368	&	5849	\\
Jelly 	&	1,939	&	56	&	5,060	\\
Jexl 	&	1,551	&	119	&	34,994	\\
Jxpath  	&	597	&	265	&	4,951	\\
MINA Sshd 	&	1,370	&		&	9,031	\\
Net 	&	2,088	&	438	&	41,340	\\
Ognl 	&	608	&	3,415	&	4,945	\\
Santuario 	&	2,697	&	1,302	&	22,398	\\
Validator 	&	1,339	&	397	&	2,050	\\
Vfs 	&	2,067	&	84	&	3,719	\\
Zookeeper 	&	411	&	1,859	&	5,023	\\
\hline
\textbf{Sum} & \textbf{77,932} & \textbf{40,890} & \textbf{1,941,508}\\ 
\hline
% \multicolumn{4}{l}{\textit{``\#SQ rules violated''} means the total number of violated rule occurrences in} \\
% \multicolumn{4}{l}{the selected projects}
\end{tabular}
\end{table}

\subsection{Data Collection}
\label{DataCollection}
The Technical Debt Dataset~\cite{LenarduzziPromise2019} contains the information of the analysis of the commits of the 33 Open Source Java projects.
The commits were analyzed using different tools such as SonarQube, PyDriller~\cite{PyDriller}, Refactoring Miner~\cite{Tsantalis2018} and the SZZ algorithm~\cite{SZZ}.

In this work, we considered the following information:
\begin{itemize}
    \item \textit{SonarQube information} collected analyzing all the commits of the 33 projects with SonarQube 7.5. From SonarQube, we considered the following information available in the dataset:
    \begin{itemize}
        \item \textit{SonarQube Rules Violations.} We considered the data from the Table ``SONAR\_ISSUES'' that includes data on each rule violated in the analyzed commits.  The complete list of rules is available online\footnote{https://rules.sonarsource.com/java} but can also be found in the file ``sonar\_rules.csv'' of the Technical Debt Dataset while the diffuseness of each rule is reported in~\cite{Saarimaki2019}.
        As reported in Table~\ref{tab:SelectedProjects}, the projects analyzed violated 174 SonarQube rules 1,914,508 times. Since in our previous work~\cite{LenarduzziJSS2020} we found incongruities in the rules type and severity assigned by SonarQube, we decided to consider all the detected rules. Table~\ref{tab:TypeSeverity} shows the SonarQube ruled violated grouped by type and severity. 
        \item \textit{Software Metrics calculated by SonarQube.} We considered the \textit{24 software metrics} measured by SonarQube (table ``SONAR\_MEASURES'' of the Technical Debt data set) as listed in Table~\ref{tab:metrics}, related to size (11 types), Complexity (5 types), test coverage (4 types), and Duplication (4 types).
    \end{itemize}
    \item \textit{Fault-inducing and Fault-fixing commits identification.} In the dataset, the fault-inducing and fault-fixing commits are determined using the SZZ algorithm~\cite{SZZ,LenarduzziOpenSZZ} and reported in the table ``SZZ\_FAULT\_INDUCING\_COMMITS''. The SZZ algorithm identifies the fault-introducing commits from a set of fault-fixing commits. The fault-introducing commits are extracted from a bug tracking system such as Jira or looking at commits that state that they are fixing an issue. A complete description of the steps adopted in the SZZ algorithm is available in ~\cite{SZZ}.
% ~\cite{SZZ,LenarduzziOpenSZZ,LenarduzziPromise2019}. 
\end{itemize}

\begin{table}[]
\centering
\caption{Type and Severity of SonarQube rules violated in our projects}
\label{tab:TypeSeverity}
\begin{tabular}{p{1.2cm}|p{1.7cm}|p{0.6cm}|p{1.8cm}} \hline 
\multicolumn{2}{c|}{\textbf{SonarQube rules}} & \textbf{\#} & \textbf{occurrences} \\ \hline
\multirow{3}{*}{\textbf{Type}}  &  Bugs &   37 & 22,620\\
       & Code Smells & 130 &  1,861,999\\
       & Vulnerability & 7 & 57,489\\ \hline
\multirow{5}{*}{\textbf{Severity}} & Blocker & 8 & 18,083\\
        & Critical & 42 & 143,293\\
        & Major & 90 & 983,647\\
        & Minor & 32 & 727,155\\
        & Info & 2 & 69,330\\ \hline
\end{tabular}
\end{table}

\begin{table} []
\footnotesize
\centering
\caption{The 24 software metrics detected by SonarQube} 
\label{tab:metrics} 
\begin{tabular}
{@{}p{1.1cm}|p{10.3cm}@{}}
\hline 
\textbf{Metric} & \textbf{Description}		\\	\hline
\multicolumn{2}{c}{\textbf{Size}}	\\	\hline
NC 	&	Number of classes (including nested classes, interfaces, enums and annotations).	\\	
NF 	&	Number of files.	\\	
LL 	&	Number of physical lines (number of carriage returns).	\\	
NCLOC 	&	Also known as Effective Lines of Code (eLOC). Number of physical lines that contain at least one character which is neither a whitespace nor a tabulation nor part of a comment. 	\\	
% NCLOC-LD 	&	Non Commenting Lines of Code Distributed By Language	\\	
NCI 	&	Number of Java classes and Java interfaces 	\\	
MPI 	&	Missing package-info.java file (used to generate package-level documentation) 	\\	
P 	&	Number of packages	\\	
STT 	&	Number of statements.	\\	
% ND	&	Number of directories in the project, also including directories not containing code (e.g., images, other files...).	\\	
NOF 	&	Number of functions. Depending on the language, a function is either a function or a method or a paragraph.	\\	
NOC 	&	Number of lines containing either comment or commented-out code.
Non-significant comment lines (empty comment lines, comment lines containing only special characters, etc.) do not increase the number of comment lines.''	\\	
NOCD 	&	Density of comment lines = Comment lines / (Lines of code + Comment lines) * 100	\\	\hline
\multicolumn{2}{c}{\textbf{Complexity}}			\\	\hline
COM	&	It is the Cyclomatic Complexity calculated based on the number of paths through the code. Whenever the control flow of a function splits, the complexity counter gets incremented by one. Each function has a minimum complexity of 1. This calculation varies slightly by language because keywords and functionalities do.	\\	
CCOM 	&	Complexity average by class	\\	
FC 	&	Complexity average by method	\\	
% FCOMD 	&	Distribution of method complexity	\\	
% FCD 	&	Distribution of complexity per class	\\
COGC 	&	How hard it is to understand the code's control flow.	\\	
PDC	&	Number of package dependency  cycles 	\\	\hline
\multicolumn{2}{c}{\textbf{Test coverage}}	\\	\hline
COV 	&	It is a mix of Line coverage and Condition coverage. Its goal is to provide an even more accurate answer to the following question: How much of the source code has been covered by the unit tests?	\\	
LTC 	&	Number of lines of code which could be covered by unit tests (for example, blank lines or full comments lines are not considered as lines to cover).	\\	
LC 	&	On a given line of code, Line coverage simply answers the following question: Has this line of code been executed during the execution of the unit tests?	\\	
UL 	&	Number of lines of code which are not covered by unit tests.	\\	\hline
\multicolumn{2}{c}{\textbf{Duplication}}		\\	\hline
DL 	&	Number of lines involved in duplications	\\	
DB 	&	Number of duplicated blocks of lines.	\\	
DF 	&	Number of files involved in duplications.	\\	
DLD 	&	= (duplicated lines $\div$ lines) * 100	\\	\hline
\end{tabular}
\end{table}

\subsection{Data Analysis}
\label{sec:DataAnalysis}
% \todo[inline]{@Francesco and @Sergio}
In this Section, we report the data analysis protocol adopted in this study including data preprocessing, data analysis, and accuracy comparison metrics. 

\subsubsection{Data Preprocessing}
% \todo[inline]{per francesco: le tue RQ di riferimento sono RQ1.1. per Machine e RQ1.2 per DL}

In order to investigate our RQs we need to preprocess the data available in the Technical Debt Dataset. Moreover, since we are planning to adopt machine learning and Deep Learning techniques, we need to preprocess the data accordingly to the models we are aim to adopt. 

The preprocessing was composed by three steps: 
\begin{itemize}
    \item Data extraction from the Technical Debt Dataset
    \item Data preparation for the Machine Learning Analysis
    \item Data preparation for the Deep Learning Analysis
\end{itemize}

\begin{figure} [H]
\centering
\includegraphics[width=0.9\linewidth]{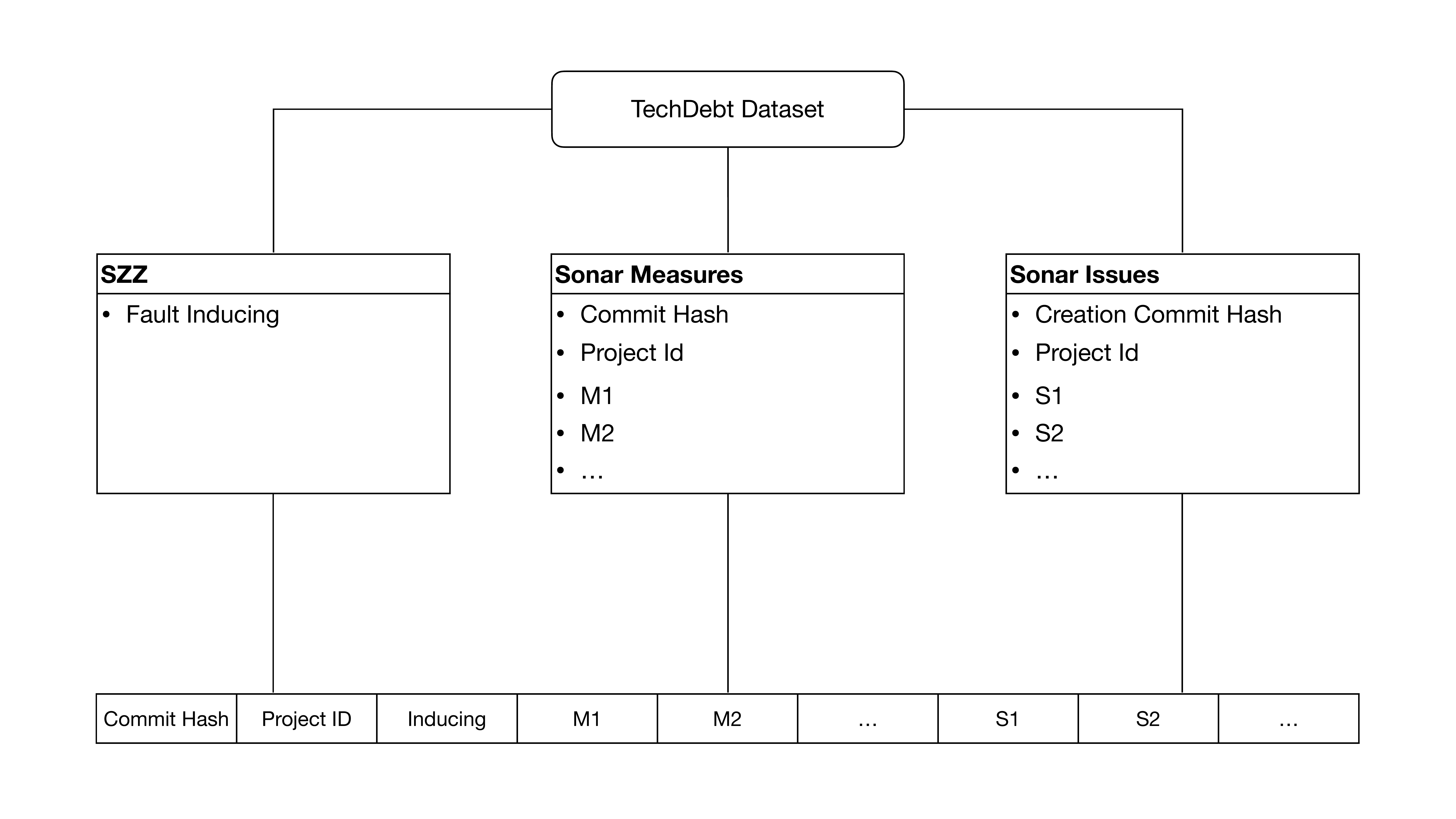}
\caption{The Data Preprocessing Process}
\label{fig:pre-process}
\end{figure}

\noindent\textbf{Data extraction from the Technical Debt Dataset.} The data in the tables SZZ\_FAULT\_INDUCING\_COMMITS, and SONAR\_MEASURES of the Technical Debt Dataset already list the information per commit (Figure~\ref{fig:pre-process}. However, the table SONAR\_ISSUES contains one row for each file where a rule has been violated. Therefore, we extracted a new table by means of an SQL query (see the replication package for details~\cite{ReplicationPackage}). The result is the new table SONAR\_ISSUE\_PER\_COMMIT. 
Then, we joined the newly created table SONAR\_ISSUE\_PER\_COMMIT with the tables SZZ\_FAULT\_INDUCING\_COMMITS, and SONAR\_MEASURES using the commit hash as key. 
This last step resulted in the final dataset that we used for our analysis (Table FullTable.csv in the replication package~\cite{ReplicationPackage}), which contains the following information: the commit hash, the project to which the commit refers to, the boolean label \textit{Inducing}, which indicates if the commit is fault inducing or not, and the set of sonar measures and sonar issues introduced in the commit. 
The complete process is depicted in Figure~\ref{fig:pre-process}.

% \begin{figure} [H]
% \centering
% \includegraphics[width=\linewidth]{./Figures/ProcessLinear.pdf}
% \caption{The Data Analysis Process}
% \label{fig:process}
% \end{figure}

% \begin{itemize}
% \item We considered 3 tables of the dataset: ISSUE, MEASURE, SZZ
% \item We extracted the commit fault-inducing. for each commit we join the information of ISSUE and MEASURE.
% \item ...
% \item Analyze the fault-inducing commits with DL
% \item Analyze the fault-inducing commits with ML
% \item Comparison DL-ML
% \end{itemize}

\vspace{2mm}
\textbf{Data preparation for the Machine Learning Analysis}. In order to predict if a commit is fault-inducing or not, based on the violation of a SonarQube rule, or to the change of a metric, we identified the fault inducing (boolean) variable as target (dependent) variable. 
% we are talking about a classification problem, our target variable \textit{y} is a binary variable indicating whether the commit is fault inducing or not. Regarding the features we used for the classification, we subdivided the problems into two steps:
% we run the analysis using respectively only the SonarQube violations and the software metrics described in section~\ref{DataCollection}.

% The structure of the data used for RQ$_{1.1}$ and RQ$_{1.2}$ is slightly different. 
The machine learning models described in section~\ref{sec:MachineLearning}, allow only to have a two dimensional input \textit{NxM}, where \textit{N} is the number of samples and \textit{M} is the number of features. This means that we can possibly classify a commit as fault inducing or not, only based on the information related to that commit itself: we cannot include the history of the commit.
For this reason, to prepare the data for answering both RQs, for each commit we selected the target variable, which is the boolean label \textit{Inducing}, and as features we prepare one input in which we use the Sonar Issues as features (RQ$_{1}$), and one in which we use the Sonar Measures (RQ$_{2}$). For the first we have a total of 77929 samples, each with 161 features each (77929x161), while for the latter we have the same number of samples and 24 features (77929x24). 

The samples are shuffled and divided in train and test set. It is important to notice that at this point, we are interested in classifying a snapshot of the commit as fault inducing or not, therefore the time dependency information is not taken into account.

%This process can be seen in described in Figure~\ref{fig:MLpreprocessing}.

% \begin{figure} [H]
% \centering
% \includegraphics[width=0.9\linewidth]{./Figures/Saner2.pdf}
% \caption{The Machine Learning preprocessing (RQ$_{1.1}$)}
% \label{fig:MLpreprocessing}
% \end{figure}

\begin{figure} [H]
\centering
\includegraphics[width=0.9\linewidth]{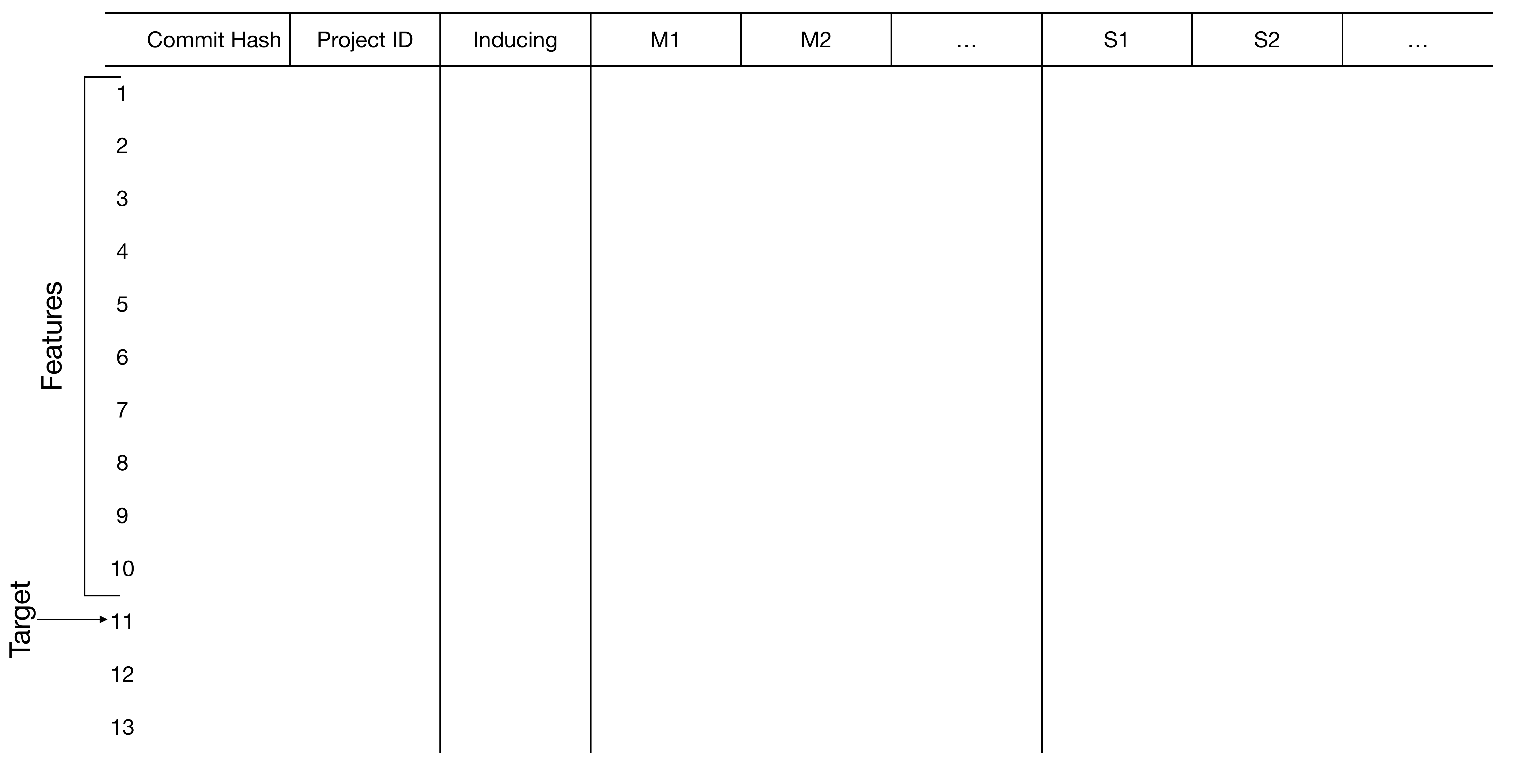}
\caption{The Deep Learning preprocessing - (RQ$_{1.2}$)}
\label{fig:DLpreprocessing}
\end{figure}

\vspace{2mm}
\textbf{Data preparation for the Deep Learning Analysis}. The deep learning models (RQ$_{1.2}$) described in Section~\ref{sec:DeepLearning}, allow the use of three dimensional input \textit{NxhxM}, where \textit{N} and \textit{M} are the number of samples and number of features, as for the machine learning models, while \textit{h} indicates the number of commits in each sample. This means that we are able to include the features related to the past commits in the classification of another commit (Figure~\ref{fig:DLpreprocessing}): we can include the history of the commit, and are not limited at using only its current status.
For this reason, we had to reshape the data in order to include the past status of the commits. We used the previous 10 commits as input variable for our models, and the label of the following commit as target variable.
Going more in details, as we have multiple projects in our dataset, we firstly divided the data into subsets each including only one project. This simply helps us in including only commits from the same project in each sample. After doing this, we reshaped the data using a rolling window of length 10 and step 1, selecting 10 commits and storing the following commit label as target variable. We did this iteratively for all the commits for each project. Also, as for the machine learning case, we prepared two different input files, one in which we use Sonar Issues as input (RQ$_{1}$) and one in which we used the Sonar Measures (RQ$_{2}$). After this step, the dimension of the input data is the following: for Sonar Issues we have the same number of sample, including 10 commits and 161 features each (77632x10x161); for the Sonar Measures we have an input of 77632 samples, including 10 commits and 24 features each (77632x10x24). This whole process can be seen in Figure~\ref{fig:DLpreprocessing}. 

Once the new samples are obtained, they are shuffled and divided in train and test set. Contrary to the machine learning case, here we take into account the time dependency between commits, but it is indeed important to notice that this is done in each individual sample, and therefore it is not necessary to consider any temporal order in the train-test split.

% \todo[inline]{@Francesco: da qui in poi si parla dell'applicazione dei processi di Deep learning, non di preprocessing. QUesta parte va spostata! }
% \todo[inline]{@Vale: hai ragione. Rimuovo la parte su 10-folds perché c'è già nella parte di "Accuracy Comparison".
% La spiegazione su feature importance la sposto in una sottosezione dopo "Accuracy Comparison"}
% Once the data has been prepared, we trained the machine learning and deep learning models. We validated the results using a 10-folds stratified cross validation.

\subsubsection{Data Analysis}
We first analyzed the fault-proneness of SonarQube rule violations (RQ$_{1}$) and (RQ$_{1.1}$) and of software metrics (RQ$_{2.1}$ and RQ$_{2.1}$) with the three Machine Learning models that better performed on this task in our previous work~\cite{LenarduzziSANER2020}. Then, we applied Deep Learning models on the same data to get better insights of the data with more advanced analysis techniques. 
Finally we compared the accuracy of the results obtained. 

\vspace{2mm}
\noindent\textbf{Machine Learning Analysis (all RQs)}.
The three machine learning models presented in section~\ref{sec:MachineLearning}, were all implemented using \textit{Scikit-learn} library, except for the XGBoost model, implemented using its own library. All the classifiers were trained using 100 decision trees. The models were trained using 80\% of the data for training and 20\% for validating the models. All three ML models were run on an Intel Xeon W-2145 with 16 cores and 64GB of RAM. The execution time for the models based on the feature set used is shown in Table~\ref{tab:ExecutionTime}

\vspace{2mm}
\noindent\textbf{Deep Learning Analysis (all RQs)}.
% \todo[inline]{@Francesco: Appena finiscono i risultati, scrivere qui i dettagli sul training (hyperparameters, tempo, epochs, ecc.)}
The deep learning models described in section~\ref{sec:DeepLearning}, were implemented in TensorFlow~\cite{tensorflow2015-whitepaper} and Keras~\cite{chollet2015keras}, using a similar approach as~\cite{fawaz2019deep}. Both models were trained for 500 epochs, with a mini-batch size of 64 and using as optimizer the Adadelta algorithm~\cite{zeiler2012adadelta}, which allow the model to adapt the learning rate.
In order to better compare the results with the ones obtained using classical machine learning methods, also the deep learning models were trained on 80\% of the data and validated using the remaining 20\%. Both models have been trained on a computational cluster with a total of 32 NVIDIA Tesla P100 and 160 CPU cores specific for training deep learning models. Each of our model had available 1 NVIDIA Tesla P100 with 16GB of VRAM, 1 CPU core, and 40GB of RAM.

\begin{table}[!ht]
\centering
\caption{Execution Times of each model based on the feature used}
\label{tab:ExecutionTime}
\begin{tabular}{c|c|c}
\hline
\textbf{Model} & \textbf{SonarQube Rules (RQ$_{1}$)} & \textbf{Software Metrics (RQ$_{2}$)} \\
\hline
Random Forest & 171s & 32s\\ 
Gradient Boosting & 353s & 159s\\ 
XGBoost & 113s & 33s\\ 
FCNN & 14h 56m & 13h 39m\\ 
ResNet & 24h 40m & 20h 53m\\ 
\hline

\end{tabular}
\end{table}

\vspace{2mm}
\noindent\textbf{Accuracy Comparison (all RQs)}.
\label{sec:accuracy}
To assess the prediction accuracy (for all the RQs), we performed a 10-fold stratified cross-validation, dividing the data in 10 parts, \textit{i.e.,} we trained the models ten times always using 1/10 of the data as a testing set. For each fold, we evaluated the classifiers by calculating a number of accuracy metrics (see below). The data has been split in a stratified manner, therefore maintaining the proportion between classes in each fold. 

As accuracy metrics, we first calculated precision and recall. However, as suggested by~\cite{Powers2011}, 
these two measures present some biases as they are mainly focused on positive examples and predictions and they do not capture any information about the rates and kind of errors made. 

The contingency matrix (also named confusion matrix), and the related f-measure help to overcome this issue. Moreover, as recommended by~\cite{Powers2011}, the Matthews Correlation Coefficient (MCC) should be also considered to understand possible disagreement between actual values and predictions as it involves all the four quadrants of the contingency matrix.
From the contingency matrix, we retrieved the measure of \textit{true negative rate} (TNR), which measures the percentage of negative sample correctly categorized as negative, \textit{false positive rate} (FPR) which measures the percentage of negative sample misclassified as positive, and \textit{false negative rate} (FNR), measuring the percentage of positive samples misclassified as negative. The measure of \textit{true positive rate} is left out as equivalent to the recall.
The way these measures were calculated can be found in Table \ref{tab:formula}.

\begin{table}[!ht]
\centering
\caption{Accuracy Metrics Formulae}
\label{tab:formula}
\begin{tabular}{c|c}
\hline
\textbf{Accuracy Measure} & \textbf{Formula}\\
\hline
Precision &\(\frac{TP}{FP + TP}\) \\ 
Recall &\(\frac{TP}{FN + TP}\)\\ 
MCC &\(\frac{TP * TN - FP * FN}{\sqrt{(FP + TP)(FN + TP)(FP + TN)(FN + TN)}}\)\\ 
f-measure &\(2* \frac{precision * recall}{precision + recall}\)\\ 
TNR &\(\frac{TN}{FP + TNe}\)\\ 
FPR &\(\frac{FP}{TN + FP}\)\\ 
FNR &\(\frac{FN}{FN + TP}\)\\ \hline

\end{tabular}
\smallskip

TP: True Positive; TN: True Negative; FP: False Positive; FN: False Negative
\vspace{-4mm}
\end{table}

Finally, to graphically compare the true positive  and the false positive rates, 
we calculated the Receiver Operating Characteristics (ROC), and the related Area Under the Receiver Operating Characteristic Curve (AUC). This gives us the probability that a classifier will rank a randomly chosen positive instance higher than a randomly chosen negative one.

In our dataset, the proportion of the two types of commits is not even: a large majority (approx. 99\%) of the commits were non-fault-inducing, and a plain accuracy would reach high values simply by always predicting the majority class. On the other hand, the ROC curve (as well as the precision and recall scores) are informative even in seriously unbalanced situations.

\vspace{2mm}
\noindent\textbf{Feature Importance (RQ$_{1.1}$ and RQ$_{2.1}$)}.
\label{sec:feature_importance}
In this work, we also studied what is the impact of each individual feature (sonar issue or measure) on the classification accuracy of the machine learning and deep learning models used. To do this, we used a \textit{permutation feature importance} technique: we first identified the baseline accuracy of the models (expressed in terms of Area Under the Curve (AUC). After this, we iteratively permuted one feature at the time, and tested the model on the newly obtained feature set.
This process was repeated until all the features were permuted, one at the time.
The difference in accuracy given between each new iteration, gives us the importance of the feature. 
It is important to notice that this feature importance method gives the importance related to only one feature at the time compared to all the remaining features. We can't therefore assume that removing all the worst performing features will increase the classification accuracy of the models. Moreover, as the importance obtained is only as good as the classification performance of the model used, the features were ranked based on the importances calculated by the best performing model.

\subsection{Replicability}
\label{sec:Replicability}
In order to allow the replication of our study, we published the complete raw data, including all the scripts adopted to perform the analysis and all the results in the replication package~\cite{ReplicationPackage}.

% \todo[inline]{spostare replication package nelle references. non mi piace come footnote}

\section{Results}
\label{sec:Results}
In this Section we first report a summary of the data analyzed and then we answer our RQs. 

We considered 77,932 commits in 33 Java projects that violated 174 different rules a total of 1,941,508 times. Out of 174 rules detected in our projects, only 161 are categorized with a SonarQube ID, and these are the ones that we used as input for our analysis, as described in section~\ref{sec:DataAnalysis}. The 455 commits labelled by SZZ as fault-inducing, violated 149 Sonarqube rules 397,595 times, as reported in Table~\ref{tab:OverviewSQ}.
Table~\ref{tab:SelectedProjects} reports the list of projects together with the number of analyzed commits while in the Table~\ref{tab:SQViolation} we report the top-20 violated SonarQube rules in the fault-inducing commits. 

In the remainder of this Section, we refer to the SonarQube Violations only with their SonarQube id number (e.g. S108). The complete list of rules, together with their description is reported in the online replication package (file  SonarQube-rules.xlsx in the replication package~\cite{ReplicationPackage}). 
% \todo[inline]{ricordatevi di aggiongewre il gfile sonar-rules ai raw data!}

\begin{table}[]
\centering
\caption{Overview of the SonarQube rules violated in the fault-inducing commits}
\label{tab:OverviewSQ}
\begin{tabular}{p{2.5cm}|p{2cm}|p{0.6cm}|p{1.8cm}} \hline 
\multicolumn{2}{c|}{\textbf{SonarQube rules in the fault-inducing commits}} & \textbf{\#} & \textbf{Occurrences} \\ \hline
\multirow{3}{*}{\textbf{Type}}  &  Bugs & 26 & 4,491 \\
& Code Smells & 116 & 374,106\\
& Vulnerability & 7 & 18,998\\ \hline
\multirow{5}{*}{\textbf{Severity}} & Blocker & 6 & 7,959 \\
& Critical & 31 & 28,647\\
& Major & 81 & 216,655\\
 & Minor & 29 & 125,993 \\
& Info & 2 & 18,341\\ \hline
\end{tabular}
\end{table}

\begin{table}[]
\centering
\caption{The top-20 violated SonarQube rules}
\label{tab:SQViolation}
\begin{tabular}{l|r|l|l} \hline 
\textbf{SonarQube rules} & \textbf{Occurrences} & \textbf{Type} & \textbf{Severity} \\ \hline 
 S134	&	23,192	&	Code Smells	&	Major	\\
 S00112	&	22,185	&	Code Smells	&	Major	\\
 RTDC	&	17,324	&	Code Smells	&	Minor	\\
 S1166	&	16,164	&	Code Smells	&	Critical	\\
 S1192	&	15,827	&	Code Smells	&	Minor	\\
 S1213	&	15,615	&	Code Smells	&	Minor	\\
 S1133	&	15,236	&	Code Smells	&	Info	\\
 S106	&	14,196	&	Code Smells	&	Major	\\
 S1132	&	13,815	&	Code Smells	&	Major	\\
 MCC	&	13,447	&	Code Smells	&	Major	\\
 MOC	&	12,533	&	Code Smells	&	Minor	\\
 S1197	&	11,861	&	Code Smells	&	Minor	\\
 COC	&	11,652	&	Code Smells	&	Major	\\
 S1312	&	11,320	&	Code Smells	&	Minor	\\
 S00117	&	11,274	&	Code Smells	&	Minor	\\
 MDC	&	10,799	&	Code Smells	&	Major	\\
 CVVC	&	10,215	&	Vulnerability	&	Major	\\ \hline 
 \multicolumn{4}{l}{RTDC means RedundantThrowsDeclarationCheck}\\
 \multicolumn{4}{l}{MCC means ``MethodCyclomaticComplexity''} \\
 \multicolumn{4}{l}{MOC means ``ModifiersOrderCheck''}\\
 \multicolumn{4}{l}{COC means ``CommentedOutCodeLine''}\\
 \multicolumn{4}{l}{MDC means ``MissingDeprecatedCheck''}\\
 \multicolumn{4}{l}{CVVC means ``ClassVariableVisibilityCheck''} \\
\end{tabular}
\end{table}

% Not all the rules detected in our projects, both in qualitative and quantitative terms, have been identified in the fault-inducing commits. 

\begin{figure} [H]
    \centering
     \includegraphics[width=0.9\linewidth, trim={10 10 5 10}, clip]{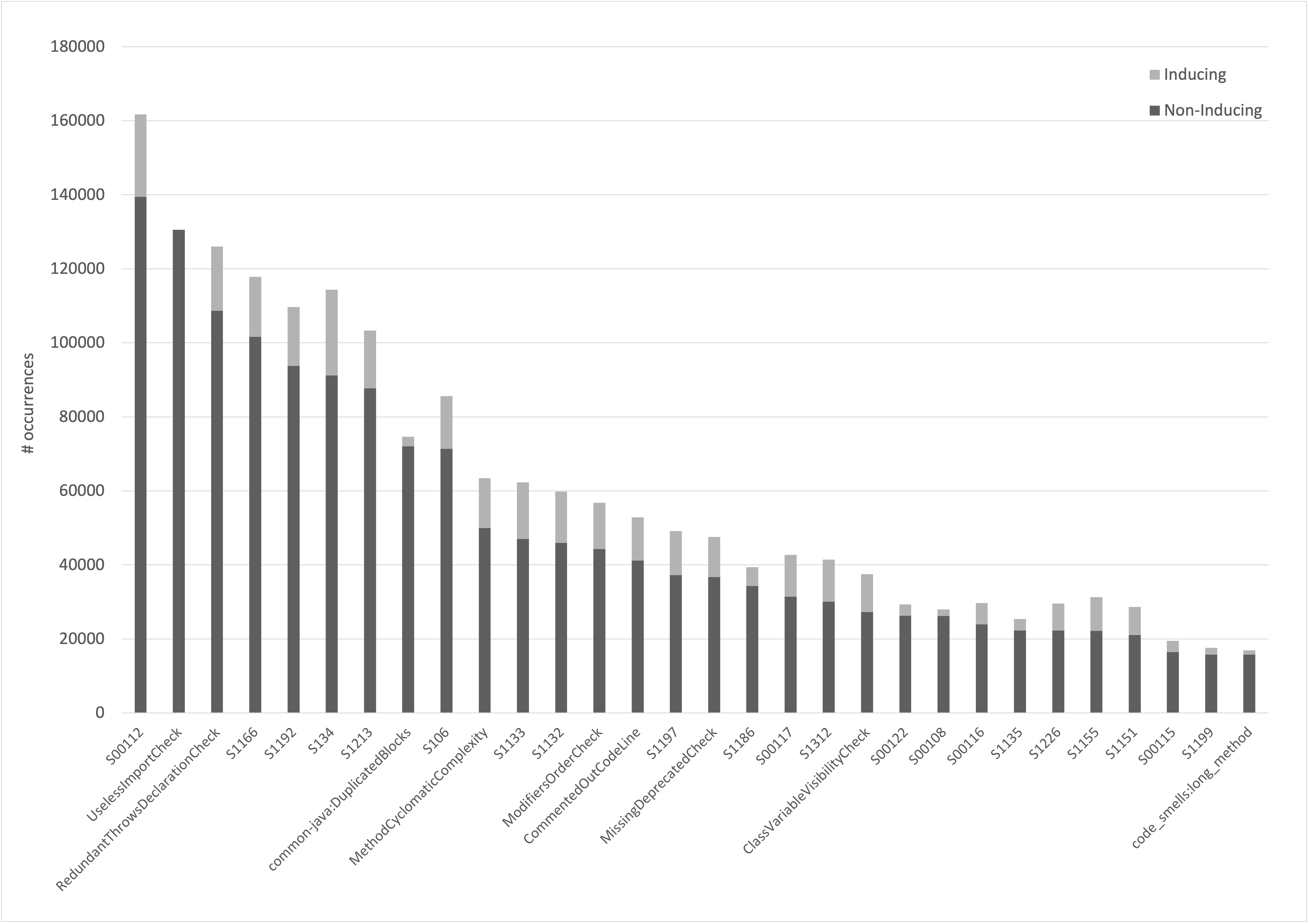}
    \caption{Distribution of the top 30 SonarQube violations detected the inducing and not-inducing commits}
    \label{fig:Diffusion}
\end{figure}

It is important to remember that, according to the SonarQube model a Bug ``represents something wrong in the code and will soon be reflected in a fault''. Moreover, they also claim that zero false positives are expected from bugs\footnote{\label{sq-rules}SonarQube Rules: https://tinyurl.com/v7r8rqo}. Therefore, we should expect that Bugs represented the vast majority of the rules detected in the fault-inducing commits. 
However, all the three types present a similar distribution: 19.85\% of Bug, 20.09\% of Code Smells, and 33.04\% of Security Vulnerabilities. Figure~\ref{fig:Diffusion} depicts the distribution of the top 30 SonarQube violations detected the inducing and not-inducing commits. 
% This confirms the results achieved in our previous studies~\cite{LenarduzziSANER2020, LenarduzziJSS2020} where we found some critical issues related to the impact of the bug rules to a presence of a fault. 

\subsection{RQ$_1$. What is the fault proneness of all the SonarQube rules?}

We analyzed our projects with the three selected machine earning (ML) models (Gradient Boost, Random Forest, and XG Boost) and with two deep learning (DL) models (FCNN and RN) to predict a fault based on SonarQube rules. 

% Table~\ref{tab:AccuracySQViolationRQ1} reports the results for the three selected Machine Learning models. 

\begin{figure} []
\centering
     \includegraphics[width=0.65\textwidth, clip]{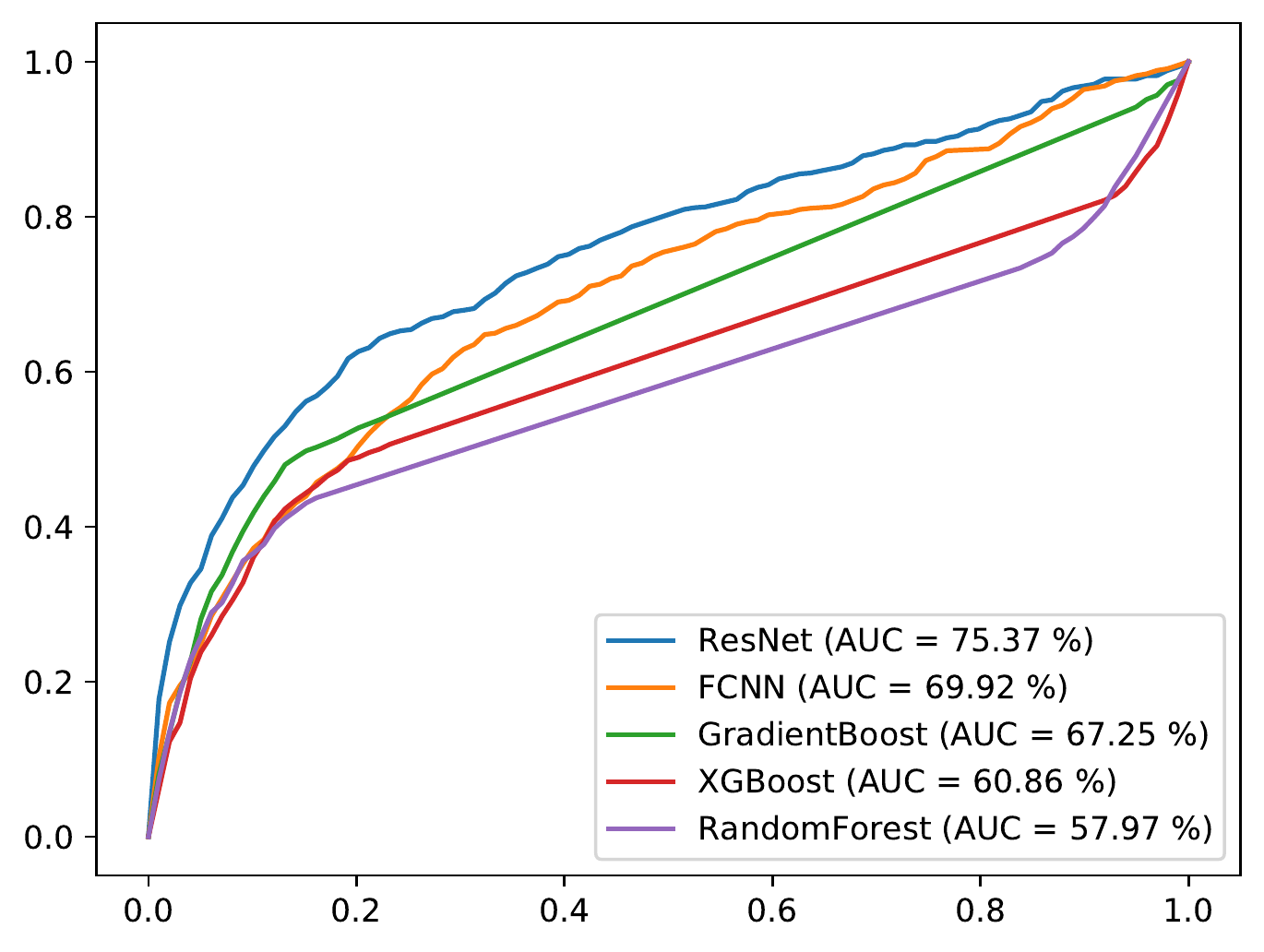}
    \caption{Area under the curve (AUC) comparison (RQ$_1$)}
    \label{fig:AccuracyRQ1}
\end{figure}

For the three Machine Learning models, the 10-fold cross-validation reported an average AUC of ~60\% (as also shown in Figure~\ref{fig:AccuracyRQ1} and in Table~\ref{tab:AccuracySQViolationRQ1}. 

As in our previous work~\cite{LenarduzziSANER2020}, the results are consistent between all the models, showing a big discrepancy in magnitude for the different accuracy metrics used. We focused our attention on the area under the curve (AUC), calculated for the receiver operating characteristic (ROC) curve. This measures gives us an overall information on how well the model can discriminate between the two classes, and therefore it helps us at comparing the classifiers one against the other. The other measures, on the other hand, give us precise information on the classification itself. It can be seen that all the measures beside the AUC are poor, and it can be especially noted an extremely high false negative rate (FNR) and true negative rate (TNR): both of them are ~99\% for all the three ML models. From this, we can deduce that the models are learning to predict the class \textit{0}, i.e. negative, which yield a 99\% accuracy rate on the negative class, but perform very poorly at detecting the positive class. This is due to the specific dataset that we are working with: we have in fact less than 1\% of the commits which are fault inducing, and therefore it becomes very difficult for a binary classifier to properly learn the features of both the negative and positive class.
It is important to note that this analysis has been performed considering only the number of rules violated in the current commit, and did not consider the historical variation of the data.  

\begin{table}[]
\centering
\caption{Fault prediction based on SonarQube rules - Accuracy Metrics Comparison (RQ$_1$)}
\label{tab:AccuracySQViolationRQ1}
\begin{tabular}{l|r|r|r|r|r}
\hline
\multirow{2}{*}{\textbf{SonarQube rules}} & \multicolumn{3}{c|}{\textbf{Machine Learning}} & \multicolumn{2}{c}{\textbf{Deep Learning}} \\ \cline{2-6}
& Gradient Boost & Random Forest & XG Boost & FCNN & RN\\ \hline
AUC & 67.25 & 57.97 & 60.86 & 69.92 & 75.37\\
Precision  & 2.64 & 1.11 & 0.59 & 4.87 & 6.79\\ 
Recall  & 2.61 & 0.44 & 0.22 & 17.22 & 24.19\\ 
MCC  & 1.61 & 0.58 & 0.27 & 7.35 & 10.68\\ 
f-measure  & 1.41 & 0.63 & 0.32  & 6.95 & 9.21\\ 
TNR  & 99.71 & 99.96 & 99.97 & 96.89 & 97.17\\ 
FPR & 0.28 & 0.04 & 0.03 & 3.11 & 2.83\\ 
FNR & 96.39 & 99.57 & 99.78 & 82.78 & 75.81\\ \hline
\end{tabular}
\end{table}

% \begin{table}[H]
% \centering
% \begin{tabular}{ll}

% \begin{minipage}{0.5\textwidth}
%     \includegraphics[width=\linewidth]{Figures/XGBoost_squid_full.pdf}
%     \captionof{figure}{Accuracy Model (AUC) - Gradient Boost (RQ$_1$)} 
%     \label{fig:AccuracyGB}
% \end{minipage}

% &

% \begin{minipage}{0.5\textwidth}
% \includegraphics[width=\linewidth, clip]{Figures/RandomForest_squid_full}
%     \captionof{figure}{Accuracy Model (AUC) - Ramdom Forest (RQ$_1$)} 
%     \label{fig:AccuracyRF}
% \end{minipage}

% \end{tabular}
% \end{table}

% \begin{figure} [H]
%      \includegraphics[width=0.5\textwidth, clip]{Figures/GradientBoost_squid_full}
%     \caption{Accuracy Model (AUC) - XG Boost (RQ$_1$)}
%     \label{fig:AccuracyXG}
% \end{figure}

In order to consider the impact of the trends of our data, we applied the two Deep Learning models. Results are much more accurate (Table~\ref{tab:AccuracySQViolationRQ1}). However, considering all the accuracy measures (Table~\ref{tab:AccuracySQViolationRQ1}), we have a similar situation to the one described for the Machine Learning models. In this case we can see that in terms of AUC both the deep learning models outperformed the machine learning model, with an AUC of ~70\% for the FCNN and ~75\% for the ResNet. For the other accuracy metrics, we have instead again poor results (although better than with the machine learning models). We can notice from these results the same trend of poor performance in terms of false negative rate (FNR), highlighted by the very high true negative rate (TNR).

% \begin{table}[H]
% \centering
% \caption{Fault prediction based on SonarQube rules - Accuracy Metrics (RQ$_1$)}
% \label{tab:AccuracySQViolationDLRQ1}
% \begin{tabular}{l|r|r}
% \hline
% \textbf{Accuracy} & \textbf{FCNN} & \textbf{RN} \\
% \hline
% AUC & 70.75 & 76.81\\
% Precision & 4.64 & 9.40\\ 
% Recall & 16.35 & 20.63\\ 
% MCC & 6.68 & 12.08\\ 
% f-measure & 6.23 & 11.81\\ 
% TNR & 96.88 & 97.05\\ 
% FPR & 3.12 & 2.50\\ 
% FNR & 83.65 & 79.37\\ \hline
% \end{tabular}
% \end{table}

% \begin{table} [H]
% \centering
% \begin{tabular}{ll}

% \begin{minipage}{0.5\textwidth}
% \includegraphics[width=\linewidth, clip]{Figures/fcn_squid.pdf}
%     \captionof{figure}{Accuracy Model (AUC)-FCNN (RQ$_1$)} 
%     \label{fig:AccuracyFCNN}
% \end{minipage}

% &

% \begin{minipage}{0.5\textwidth}
%     \includegraphics[width=\linewidth]{Figures/resnet_squid.pdf}
%     \captionof{figure}{Accuracy Model (AUC)-RN (RQ$_1$)} 
%     \label{fig:AccuracyRN}
% \end{minipage}

% \end{tabular}
% \end{table}

% \begin{figure} [H]
% \centering
% \includegraphics[width=0.65\linewidth]{Figures/fcn_squid.pdf}
%     \caption {Mean AUC with standard deviation - FCNN (RQ$_1$)} 
%     \label{fig:AccuracyFCNN}
% \end{figure}

% \begin{figure} [H]
% \centering
%     \includegraphics[width=0.65\linewidth]{Figures/resnet_squid.pdf}
%     \caption {Mean AUC with standard deviation - RN (RQ$_1$)} 
%     \label{fig:AccuracyRN}
% \end{figure}

\smallskip
\roundedbox{SonarQube rules a good predictors of a fault. Using historical data (Deep Learning) provide a better overall accurate fault detection accuracy compared to using a single snapshot (Machine Learning).}

\subsection{RQ$_{1.1}$. What is the fault proneness of each individual SonarQube rule?}

% The application of the three machine learning models, ... 
% \todo[inline]{@francesco: come sono state ranked? drop column o altro? eg. the application of the drop column algorithm enabled us to identify the importance of each rule... }
In order to calculate the fault proneness of each SonarQube rule, we used a permutation feature importance technique: we permuted one feature at the time (randomly shuffling the feature's values in the whole dataset) and calculate the change in AUC of the ML and DL models when classifying the data with a permuted feature, compared to the results obtained classifying the pure data. This difference allowed us to rank the features based on how much their permutation impact (increase or decrease) the classification accuracy of the models. 
As the feature importance is only as good as the classification performance of the model used to obtain them, we ranked the features using the importance given by the ResNet model, as it is the one with the highest overall accuracy metrics.
For reason of space, in Table~\ref{tab:DLFeaturesSQViolations} we reported only the SonarQube rules which had an importance of at least 1\%, while the complete results are included in our replication package. The top-10 rule are categorized as ``Code Smells'' and the first rule classified as ''Bug'' has an importance extremely low (less than 0.3\% for S2184 rule). 

Comparing with the SonarQube rules diffusion, not all the rules that exhibit higher importance are the most diffused ones in the previous commits. For example S134 has ranked as eighth most important rule despite the fact that this is the most diffused one. 

When comparing the severity of the SonarQube rules, the highest security level found in the top-10 features is Critical (S1166). The first rule associated with the highest security level (Blocker), is found at the 45\textsuperscript{th} position (S1181), with an importance of 0.27\%. The remaining rules have mainly a Major severity level. 

% Considering the historical data using Deep Learning models, ...
% \todo[inline]{@francesco: introdurre la parte di DL}
As it can be seen from Table~\ref{tab:DLFeaturesSQViolations}, both deep learning models were able to clearly identify which of the features were more informative to the classification. In the case of the violation S1192, for example, we see that by permuting it (randomly shuffle the values of this violation in the whole dataset), the deep learning models are able to classify the commits with an AUC which is lower by 7.63\% for the FCNN and 6.97\% for the ResNet, compared to the dataset with the unpermuted S1192. 
% \todo[inline]{@francesco che significa? spiegare meglio :)}
% \todo[inline]{@davide: vedi se si capisce un po' meglio ora}
On the other hand, the machine learning models, were not able to draw this clear distinction, with most of the features impacting the AUC by less than 1\%.
Moreover, as the feature importance is strictly dependant on the model used, the ranking based on the importances tells us which feature impacts the most the classification accuracy of the model. We can therefore generalize this, saying that the feature that mostly affects the accuracy of the model, are also the ones that can better help indicate if a commit is fault inducing or not. For these reasons, as the deep learning model performed overall better than the machine learning models, we can assume that the feature importance given by the two deep neural network is more accurate than the one given by the machine learning models.

\begin{table}[]
\centering
\caption{Fault prediction based on SonarQube rules - Feature importance (RQ$_{1.1}$)}
\label{tab:DLFeaturesSQViolations}
\begin{tabular}{l|l|l|r|r|r|r|r}
\hline
\textbf{SQ rules} & \textbf{Type} &  \textbf{Severity} 
& \multicolumn{1}{|p{1.2cm}|}{\centering \textbf{Gradient \\ Boost}}
& \multicolumn{1}{|p{1.2cm}|}{\centering \textbf{Random \\ Forest}}
& {\centering \textbf{XG Boost}} & \textbf{FCNN} & \textbf{RN} \\\hline
squid:S1192	&	CS	&	MINOR	&	5,41 \%	&	2,92 \%	&	3,20 \%	&	7,63 \%	&	6,97 \%	\\
squid:S00112	&	CS	&	MAJOR	&	-0,22 \%	&	-1,25 \%	&	0,03 \%	&	2,04 \%	&	3,10 \%	\\
squid:S1166	&	CS	&	CRITICAL	&	0,64 \%	&	-0,39 \%	&	0,06 \%	&	1,96 \%	&	2,35 \%	\\
squid:S1213	&	CS	&	MINOR	&	-0,02 \%	&	0,11 \%	&	0,36 \%	&	0,92 \%	&	2,35 \%	\\
squid:S1135	&	CS	&	INFO	&	0,08 \%	&	0,15 \%	&	0,28 \%	&	0,57 \%	&	2,26 \%	\\
squid:S106	&	CS	&	MAJOR	&	-0,39 \%	&	0,77 \%	&	-0,68 \%	&	1,93 \%	&	2,00 \%	\\
squid:COCL	&	CS	&	MAJOR	&	0,45 \%	&	0,25 \%	&	0,06 \%	&	0,52 \%	&	1,91 \%	\\
squid:S134	&	CS	&	MAJOR	&	2,13 \%	&	0,91 \%	&	1,40 \%	&	1,98 \%	&	1,75 \%	\\
squid:S1133	&	CS	&	INFO	&	-0,20 \%	&	-0,18 \%	&	0,67 \%	&	1,08 \%	&	1,44 \%	\\
squid:MCC	&	CS	&	MAJOR	&	-0,52 \%	&	-0,50 \%	&	0,33 \%	&	0,56 \%	&	1,29 \%	\\
squid:S00100	&	CS	&	MINOR	&	0,01 \%	&	0,15 \%	&	0,21 \%	&	0,64 \%	&	1,19 \%	\\
squid:S1186	&	CS	&	MAJOR	&	-0,61 \%	&	-0,39 \%	&	-0,44 \%	&	0,53 \%	&	1,11 \%	\\
squid:S1197	&	CS	&	MINOR	&	0,01 \%	&	-0,10 \%	&	-0,47 \%	&	0,09 \%	&	1,03 \%	\\
squid:MOC	&	CS	&	MINOR	&	-0,23 \%	&	0,16 \%	&	0,19 \%	&	1,04 \%	&	1,02 \%	\\
\hline
 \multicolumn{5}{l}{COCL means ``CommentedOutCodeLine''}\\
 \multicolumn{5}{l}{MCC means ``MethodCyclomaticComplexity''} \\
 \multicolumn{5}{l}{MOC means ``ModifiersOrderCheck''} \\
 \multicolumn{5}{l}{CS means ``Code Smells''}\\
\end{tabular}
\end{table}

\vspace{2mm}
\roundedbox{Historical data (Deep Learning)  allows a clear identification of a features set more informative to the fault identification, while a single snapshot (Machine Learning) do not provide a clear distinction.}

\roundedbox{14 SonarQube rules account for 30\% of the importance, based on historical data.}

\subsection{RQ$_2$. What is the fault proneness of all the metrics calculated by SonarQube?}

% \textbf{dividere per ML e DL}
% \textbf{The application of the Machine Learning models ... }

Similarly to RQ$_1$, we used the three selected Machine Learning models (Gradient Boost, Random Forest, and XG Boost) and with the two Deep learning models (FCNN and RN) to predict a fault based on software metrics.

Considering as predictors the software metrics calculated by SonarQube, the models accuracy is extremely low, both for the machine learning models and for the deep learning models.

\begin{figure} []
\centering
     \includegraphics[width=0.65\textwidth, clip]{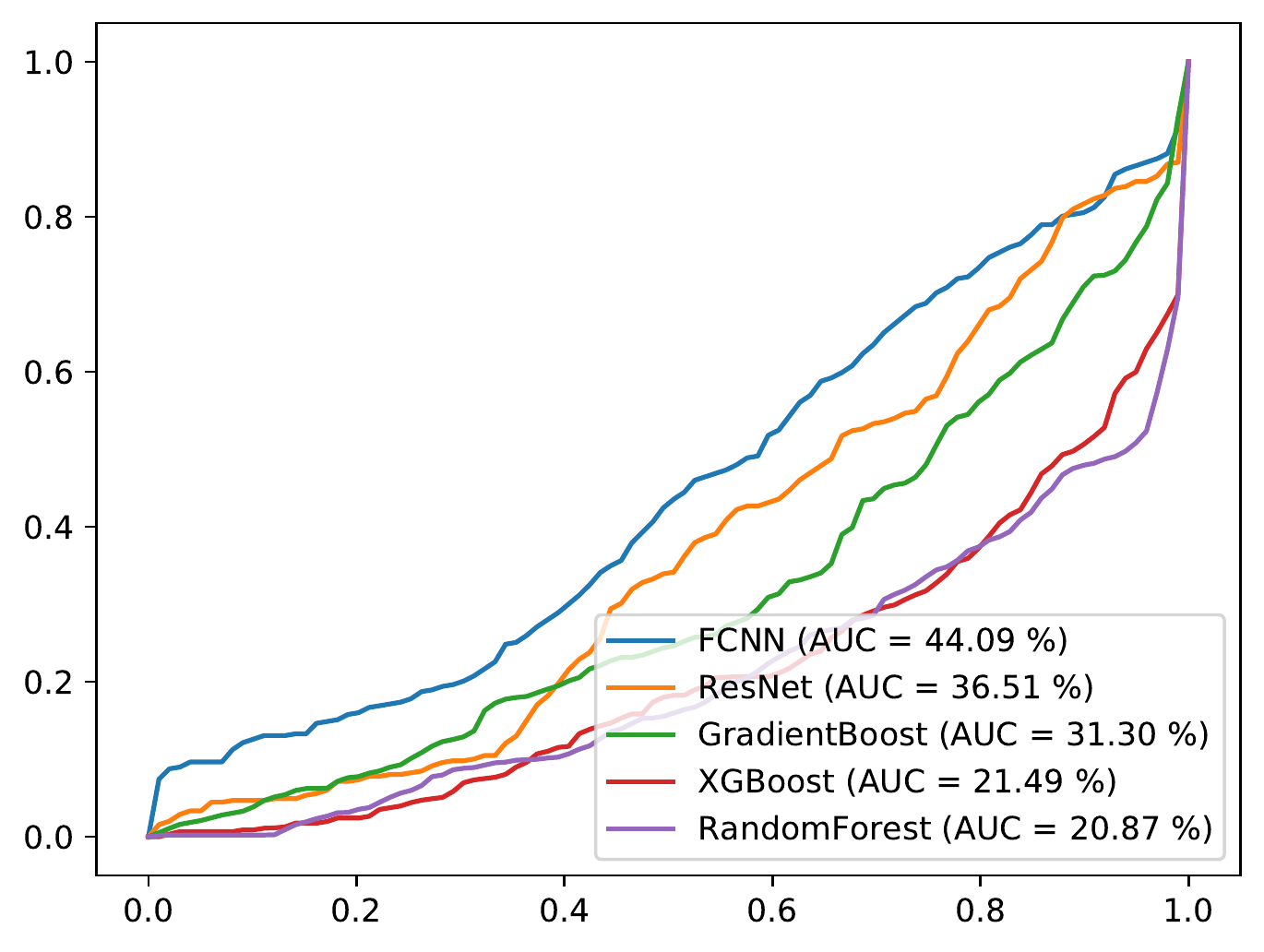}
    \caption{Area under the curve (AUC) comparison (RQ$_2$)}
    \label{fig:AccuracyML_RQ2}
\end{figure}

Going more in detail, in Figure~\ref{fig:AccuracyML_RQ2}, we can see how all three machine learning models and the two deep learning models, have a convex receiver operating characteristic (ROC) curve and hence an area under the curve lower than 50\%.

% \begin{figure} [H]
% \centering
% \includegraphics[width=0.65\linewidth]{Figures/fcn_M_.pdf}
%     \caption {Mean AUC with standard deviation - FCNN (RQ$_2$)} 
%     \label{fig:AccuracyFCNN_RQ2}
% \end{figure}

% \begin{figure} [H]
% \centering
%     \includegraphics[width=0.65\linewidth]{Figures/resnet_M_.pdf}
%     \caption {Mean AUC with standard deviation - RN (RQ$_2$)} 
%     \label{fig:AccuracyRN_RQ2}
% \end{figure}

% Similarly, we can see in Figures~\ref{fig:AccuracyML_RQ2}, that the ROC curve is closer to a straight line for the FCNN and convex for the ResNet, which in turn results in an AUC of ~50\% or less. It is also possible to notice how the standard deviation is quite high (12-13\%) for both the deep learning models, as it is also indicated by the very wide grey areas in the figures. 

\begin{table}[]
\centering
\caption{Fault prediction based on software metrics - Accuracy Metrics Comparison (RQ$_2$)}
\label{tab:AccuracyMetricsRQ2}
\begin{tabular}{l|r|r|r|r|r}
\hline
\multirow{2}{*}{\textbf{SQ metrics}} & \multicolumn{3}{c|}{\textbf{Machine Learning}} & \multicolumn{2}{c}{\textbf{Deep Learning}} \\ \cline{2-6}
& Gradient Boost & Random Forest & XG Boost & FCNN & RN\\ \hline
AUC &  31.30 & 20.87 & 21.49 & 44.09 & 36.51\\
Precision  & 0.12 & 0.03 & 0.2 & 6.12 & 10.42\\ 
Recall  & 11.76 & 1.31 & 0.87 & 52.83 & 55.54\\ 
MCC & -4.90 & -5.80 & -5.44 & -1.74 & -5.77\\ 
f-measure  & 0.23 & 0.06 & 0.03 & 2.97 & 3.19\\ 
TNR  & 59.45 & 66.36 & 66.89 & 45.56 & 32.67\\ 
FPR  & 40.55 & 33.64 & 33.11 & 54.44 & 67.33\\ 
FNR  & 88.24 & 98.69 & 99.13 & 47.16 & 44.47\\ \hline
\end{tabular}
\end{table}

Table~\ref{tab:AccuracyMetricsRQ2} reports all the accuracy metrics for the machine learning and the deep learning models. It is clear from these results that the software metrics do not appear to be a factor in determining a fault at commit level.
Since we did not obtain a good level of accuracy, we did not estimate the feature importance for the SonarQube software metrics. 

\smallskip
\roundedbox{Software metrics do not appear to be a factor in determining a fault at commit level both using historical data (Deep Learning) or a single snapshot (Machine Learning).}

% Table~\ref{tab:AccuracyMetricsRQ2} reports also the average reliability measures for the eight accuracy metrics. \todo[inline]{questa non la capisco. e' l'accuracy di cosa? del modello senza una metrica, di una singola metrica, ...?}
% Considering as predictors the software metrics calculated by SonarQube, the models accuracy is low, since the trained models are hovering around 50\%. In fact, we find an AUC hovering around 20-30\% for all models, except for the FCNN, whose AUC is ~50\% (Table~\ref{tab:AccuracyMetricsRQ2}).  Therefore, software metrics do not appear to be a factor in determining a fault at commit level. 
% Since we did not obtain a good level of accuracy, we did not estimate the feature importance for the SonarQube software metrics. 

% \textbf{The application of the Deep Learning models, ... }

% \begin{table}[H]
% \centering
% \caption{Fault prediction based on software metrics - Accuracy Metrics (RQ$_1$)}
% \label{tab:DLAccuracyMetrics}
% \begin{tabular}{l|r|r}
% \hline
% \textbf{Accuracy} & \textbf{FCNN} & \textbf{RN} \\
% \hline
% AUC & 49.89 & 35.70\\
% Precision & 6.59 & 0.46\\ 
% Recall & 58.85 & 55.12\\ 
% MCC & 11.94 & -6.75\\ 
% f-measure & 3.34 & 0.92\\ 
% TNR & 42.78 & 3.95\\ 
% FPR & 57.22 & 69.05\\ 
% FNR & 41.15 & 44.88\\ \hline
% \end{tabular}
% \end{table}

\subsection{RQ$_{2.1}$.  What is the fault proneness of each metric calculated by SonarQube?}

Since the results in RQ$_2$ are not accurate enough, the analysis of the metrics individually would only decrease the accuracy of the prediction. Therefore, we cannot proceed with the analysis of the fault-proneness of the individual metrics. 
\section{Discussion}
\label{sec:Discussion}
In this Section, we discuss the results obtained according to the RQs and present possible practical implications from our research. 

The analysis of projects contained in the Technical Debt dataset showed that Deep learning models provide a more accurate fault detection accuracy compared with machine learning ones based on SonarQube rules. Moreover, deep learning models allow a clear identification of a features set more informative to the fault identification, while machine learning models do not provide a clear distinction. 
Considering the metrics calculated by SonarQube, none of the models provide a good level of accuracy and, consequently, the features set identification. 

Out of 149 different SonarQube rules violated 397,597 times in the fault-inducing commits, we identified a rules set that can be considered as ``fault warning'' from developers (14 rules with an importance higher than 1\%, that account for 30\% of the total fault-proneness importance). 
Taking into account the SonarQube rules classification, both the most recurrent rules in the fault-inducing commits and the most important in the fault identification are ``Code Smells'' with a medium level of severity (Major, third level according SonarQube). 
This results in contract with the SonarQube model, where ``Bug'' should be the only responsible of a fault introduction in the code. We confirmed our previous works~\cite{LenarduzziSANER2020, LenarduzziJSS2020} relating to the criticism about type and severity classification rules assigned by SonarQube. 

Looking at the accuracy metrics adopted in this work, we can noticed a consistent difference in magnitude for the individual accuracy metrics. AUC provides a good level of accuracy, while Precision and Recall are both very low. This confirms our suspect that the dataset we are using, which presents ~99\% of the commits as normal commits, and the remaining 1\% as fault inducing commits, might not be suitable for a simple binary classification task, as there are not enough samples of both classes for the models to learn their characteristics. The problem might be better treated as an anomaly detection problem.

% \begin{itemize}
%     \item Sottolineare unbalance dataset
%     \item ribadire difference AUC e altre metriche
%     \item feature selection più evidente con DL
%     \item iniziare a suggerire anomaly detection ?
% \end{itemize}

 \section{Threats to Validity}
\label{sec:Threat}
In this Section, we discuss the threats to validity, including internal, external, construct validity, and reliability. We also explain the different adopted tactics~\cite{YinCaseStudies2009}. 

\smallskip
\noindent \textbf{Construct Validity}. This threat concerns the relationship between theory and observation due to possible measurement errors. 
SonarQube is one of the most adopted static analysis tool by developers~\cite{VassalloESEM2019,Avgeriou2020}. Nevertheless, we cannot exclude the presence of false positives or false negatives in the detected warnings; further analyses on these aspects are part of our future research agenda. As for code smells, we employed a manually-validated oracle, hence avoiding possible issues due to the presence of false positives and negatives. 
We relied on the ASF practice of tagging commits with the issue ID. However, in some cases, developers could have tagged a commit differently. Moreover, the results could also be biased due to detection errors of SonarQube.
We are aware that static analysis tools suffer from false positives. In this work we aimed at understanding the fault proneness of the rules adopted by the tools without modifying them, so as to reflect the real impact that developers would have while using the tools. In future works, we are planning to replicate this work manually validating a statistically significant sample of violations, to assess the impact of false positives on the achieved findings. 
As for the analysis time frame, we analyzed commits until the end of 2015, considering all the faults raised until the end of March 2018. We expect that the vast majority of the faults should have been fixed. However, it could be possible that some of these faults were still not identified and fixed.

\smallskip
\noindent \textbf{Internal Validity}. This threat concerns internal factors related to the study that might have affected the results. 
As for the identification of the fault-inducing commits, we relied on the SZZ algorithm~\cite{SZZ}.  We are aware that in some cases, the SZZ algorithm might not have identified fault-inducing commits correctly because of the limitations of the line-based diff provided by git, and also because  in some cases bugs can be fixed modifying code in other locations than in the lines that induced them. Moreover, we are aware that the imbalanced data could have influenced the results (approximately 90\% of the commits were non-fault-inducing). However, the application of solid machine learning techniques, commonly applied with imbalanced data could help to reduce this threat. 

\smallskip
\noindent \textbf{External Validity}. Our study considered the 33 Java open-source software projects with different scope and characteristics included in the Technical Debt dataset.
All the 33 Java projects are members of the Apache Software Foundations that incubates only certain systems that follow specific and strict quality rules. 
Our case study was not based only on one application domain. This was avoided since we aimed to find general mathematical models for the prediction of the number of bugs in a system. Choosing only one or a very small number of application domains could have been an indication of the non-generality of our study, as only prediction models from the selected application domain would have been chosen. The selected projects stem from a very large set of application domains, ranging from external libraries, frameworks, and web utilities to large computational infrastructures. 

The dataset only included Java projects. We are aware that different programming languages, and projects at different maturity levels could provide different results. We selected 33 projects from the ASF, which incubates only certain systems that follow specific and strict quality rules. Our case study was not based only on one application domain. This was avoided since we aimed to find general mathematical models for the prediction of the number of bugs in a system. Choosing only one or a very small number of application domains could have been an indication of the non-generality of our study, as only prediction models from the selected application domain would have been chosen. The selected projects stem from a very large set of application domains, ranging from external libraries, frameworks, and web utilities to large computational infrastructures. The dataset only included Java projects. We are aware that different  programming languages, and projects at different maturity levels could provide different results. 

\smallskip
\noindent \textbf{Conclusion Validity}. This threat concerns the relationship between the treatment and the outcome. We adopted different machine learning and deep learning models to reduce the bias of the low prediction power that a single classifier could have. We also addressed possible issues due to multi-collinearity, missing hyper-parameter configuration, and data imbalance. We recognize, however, that other statistical or machine learning techniques might have yielded similar or even better accuracies than the techniques we used.

\section{Related Work}
\label{sec:RW}
Faults prediction has been deeply investigated in the last years, where research focused mainly on improving the granularity of the predictions~\cite{Pascarella2019}, adding features, e.g., code review~\cite{McIntosh2018}, change context~\cite{Kondo2019}, or applying machine and deep learning models~\cite{Hoang2019, LenarduzziSANER2020}.

As factors to predict bug-inducing changes some authors adopted change based metrics~\cite{McIntosh2018}, including size~\cite{Kamei2013}, the history of a change as well as developer experience~\cite{Kamei2013}, or churn metrics~\cite{Tan2015}. Another study included code review metrics for the predictive models~\cite{McIntosh2018}. 
One aspect investigated was also the decreasing of the effort required to diagnose a defect~\cite{Pascarella2019}. 

Two studies included as factors static analysis warnings~\cite{Querel2018,Trautsch2020} for building just-in-time defect prediction models. According to their results, they can improve the predictive models accuracy~\cite{Querel2018}. Moreover, both code metrics and static analysis warnings are correlated with bugs and that they can improve the prediction~\cite{Trautsch2020}. 

Faults prediction were investigated adopting Machine learning models focusing on the features role such as change size or changes history, that can represent a code change, and using them as predictors~\cite{Kamei2013, Pascarella2018, Pascarella2019}. 

Machine learning techniques were also largely applied in detection of technical issues in the code, such as code smells~\cite{ArcelliFontana2016, DiNucci2018, pecorelli2020developer, LujanMaltesque2020}.
While machine learning has been mainly applied to detect different code smell types~\cite{Khomh2009, Khomh2011}, unfortunately, only few studies applied machine learning techniques to investigate static analysis tool rules, such as SonarQube~\cite{Falessi2017,Tollin2017,LenarduzziSANER2020} or PMD~\cite{LenarduzziPMD2020}.

Machine learning techniques were applied to detect multiple code smell types~\cite{ArcelliFontana2016}, estimate their harmfulness~\cite{ArcelliFontana2016}, determine the intensity~\cite{ArcelliFontana2017}, and to classify code smells according to their perceived criticality~\cite{pecorelli2020developer}. The training data selection can influence the performance of machine learning-based code smell detection approaches~\cite{DiNucci2018} since the code smells detected in the code are generally few in terms of number of occurrences~\cite{pecorelli2020large}.

Moreover, machine learning algorithms were successfully applied to derive code smells from different software metrics~\cite{Maneerat2011}. 

Considering the detection of static analysis tool rules, SonarQube was the tool mainly investigated, focusing on the effect of the presence of its rules on fault-proneness~\cite{Falessi2017, LenarduzziSANER2020} or the change-proneness~\cite{Tollin2017}. 

Machine learning approaches were successfully applied since results showed that 20\% of faults were avoidable if the SonarQube-related issues would have been removed~\cite{Falessi2017}, however, the harmfulness of the SonarQube rules is very low~\cite{LenarduzziSANER2020}. Positive results application were collected also considering class change-proneness~\cite{Tollin2017}. 

Machine learning approaches were also used to determine if the SonarQube technical debt was be predicted based also on software metrics~\cite{LenarduzziMaltesque2019}. Results demonstrated the impossibility to have positive prediction. Another point of view which has benefited from machine learning was the evaluation of the remediation effort calculated by SonarQube~\cite{SaarimakiEUROMICRO2019, BaldassarreIST2020}. Results highlighted the model overestimation of the time to fix the Technical Debt-related issues. 

In order to satisfy computer performance that are fastly increasing in the last years, Deep Learning is becoming popular in many domains~\cite{Hinton2006} such as image classification~\cite{Krizhevsky2017} or natural language processing~\cite{Sarikaya2014}. 
There also many existing studies that leverage deep learning techniques to address other problems in software engineering~\cite{White2015, Lam2015, GuFSE2016, GuoICSE2017, GuICSE2018}.
Since the promising results, Deep Learning could be a valid approach to adopt also in bug prediction in order to improve the performance of just-in-time defect prediction. 

Deep learning can be useful to improve the logistic regression weaknesses when the study should combine features to generate new ones. This approach was successfully applied in~\cite{Yang2015} considering 14 traditional change level features in order to predict bugs.

The benefit of using Deep Learning instead of machine Learning to improve the performance of just-in-time defect prediction is still under investigation~\cite{Yang2015, Abozeed2019, Ferenc2020, Wang2020}.
The results achieved until now demonstrates a promising improvement in the bug prediction accuracy compared with other approaches (32.22\% more bugs detected)~\cite{Yang2015} especially for small dataset and in the feature selection~\cite{Abozeed2019}, and to predict the presence of bugs in classes from static source code metrics~\cite{Ferenc2020}. 

Ones of the most adopted Deep Learning models to automate feature learning for defect prediction are Long Short Term Memory~\cite{Dam2021} and Convolutional Neural Network~\cite{Li2017}. Another models well-known is Deep Belief Network~\cite{Wang2020}.

\section{Conclusion}
\label{sec:Conclusions}

In this paper we investigated the fault-proneness of rules and metrics detected by SonarQube adopting Machine Learning and Deep Learning models. 

In our previous work, on a reduced dataset~\cite{LenarduzziSANER2020}, we found that SonarQube rules considered fault-inducing were not properly classified, however even if we obtained a good prediction accuracy, we were not able to accurately detect the impact of each single rule on the fault-proneness. Results were also confirmed by our next work on a extended dataset (the same considered in this work) where we applied statistical techniques to detect if the violation  of any SonarQube rule impacted the fault-proneness. 

In order to corroborate our previous results, and to clearly identify the impact of each different SonarQube rule, in this work, we applied and compared Machine Learning and Deep Learning models. Moreover, we also investigated the impact of 24 software metrics such as cyclomatic complexity and lines of code on the fault-proneness and we compared it with the fault-proneness of SonarQube rules. 

SonarQube detects more than 500 rules for Java. However, in the dataset we adopted (33 projects and 77,932 commits), only 174 were violated, and in particular, only 149 were violated in fault-inducing commits. 

As for the identification of the most important rules that affect the fault-proneness of a commit, the application of machine learning models was outperformed by the deep-learning models. The reason might lie in the data adopted for the prediction. While machine learning models were trained and tested using the data of the individual commit to be classified, the Deep Learning models were trained using the historical data coming from the previous 10 commits, hence enabling the models to extract more information from the data. 

Results show that, out of the 149 SonarQube rules detected in fault-inducing commits, fourteen of them have an importance higher than 1\%, while one rule (S1192) which has the highest importance (7\%). Therefore, developers might focus their attention only on these rules if they want to decrease the likelihood of introducing a fault in their code. 

Considering the fault-proneness of the metrics calculated by SonarQube, considering all the 24 metrics together, the fault-proneness accuracy is extremely low, and therefore the impact of each individual metric on fault-proneness is negligible. 

Future works might consider the adoption of time series analysis and anomaly detection techniques to better and more accurately detect the rules that impact fault-proneness. Another interesting avenue, is to enable a more accurate identification of the thresholds adopted by SonarQube to consider the violation of a rule. 

% \todo[inline]{@Vale: secondo me non è molto chiara quest'ultimo paragrafetto}
Moreover, several violations are defined as thresholds on existing metrics (e.g. Large Class, Long Methods, Too many Parameters, ...). As an example, SonarQube rule RSPEC-1541\footnote{SonarQube rule ``Methods should not be too complex default value'' \url{https://rules.sonarsource.com/java/RSPEC-1541}} indicates that  ``Methods should not be too complex default value''  raising a violation with the cyclomatic complexity of a method is higher than the default value 10. Therefore, it would be useful to apply slope-based thresholds~\cite{Morasca2020} or other techniques to better identify the thresholds adopted by SonarQube in its violations.

\bibliographystyle{spbasic}    
\bibliography{sample.bib}

\end{document}